\documentclass[twocolumn,tighten,times,trackchanges]{aastex631}

\usepackage{tabularx}
\usepackage{xspace}
\usepackage{array}
\usepackage{nameref}
\usepackage[utf8]{inputenc}
\usepackage[T1]{fontenc}
\usepackage{babel}
\usepackage{amsmath}
\usepackage{amssymb}
\usepackage{graphicx}
\usepackage{url}

\usepackage{hyperref}
\usepackage{orcidlink}
\usepackage{footmisc}
\usepackage{xcolor}
\usepackage{comment}
\usepackage{multirow}

\let\tablenum\relax

\usepackage[locale=US,mode=text,drop-zero-decimal,range-units=single,range-phrase=--]{siunitx}
\DeclareSIUnit\parsec{pc}
\DeclareSIUnit\Mpc{\mega\parsec}
\DeclareSIUnit\year{yr}
\usepackage{acronym}

\usepackage{amstext}

\makeatletter
\@ifpackageloaded{cleveref}{
  \AtBeginDocument{
    \def\ltx@label#1{\cref@label{#1}}%
    \def\label@in@display@noarg#1{\cref@old@label@in@display{#1}}%
    \def\label@in@mmeasure@noarg#1{%
      \begingroup%
        \measuring@false%
        \cref@old@label@in@display{#1}%
      \endgroup}%
  }
}{}
\makeatother

\protected\def\protectedacused{\acused}

\acrodef{LIGO}[LIGO]{Laser Interferometer Gravitational-Wave Observatory}
\acrodef{LHO}[LHO]{\ac{LIGO} Hanford Observatory}
\acrodef{LLO}[LLO]{\ac{LIGO} Livingston Observatory}
\acrodef{KAGRA}[KAGRA]{KAGRA}\acused{KAGRA}
\acrodef{iKAGRA}[iKAGRA]{initial \ac{KAGRA}}
\acrodef{bKAGRA}[bKAGRA]{baseline \ac{KAGRA}}
\acrodef{GEO}[GEO\,600]{GEO\,600 \ac{GW} detector}
\acrodef{aLIGO}{Advanced \ac{LIGO}}
\acrodef{A+}{Advanced+ \ac{LIGO}}
\acrodef{Asharp}[\ensuremath{\text{A}^\sharp}]{\ac{LIGO} \acs{Asharp}}
\acrodef{AdV}{Advanced \acl{Virgo}}
\acrodef{AdV+}{Advanced \acl{Virgo}+}
\acrodef{Virgo}{Virgo}\acused{Virgo}
\acrodef{VirgoNEXT}[Virgo\_nEXT]{Virgo\_nEXT}\acused{VirgoNEXT}

\acrodef{LSC}[LSC]{\acs{LIGO} Scientific Collaboration}
\acrodef{LV}[LV]{\acs{LIGO}--\acs{Virgo} Collaboration\protect\protectedacused{LVC}}
\acrodef{LVC}[LV]{\acs{LIGO}--\acs{Virgo} Collaboration\protect\protectedacused{LV}}
\acrodef{LVK}[LVK]{\acs{LIGO}--\ac{Virgo}--\ac{KAGRA} Collaboration}
\acrodef{IGWN}[IGWN]{International \ac{GWH} Observatory Network}

\acrodef{O1}[O1]{first observing run}
\acrodef{O2}[O2]{second observing run}
\acrodef{O3}[O3]{third observing run}
\acrodef{O3a}[O3a]{first half of the third observing run}
\acrodef{O3b}[O3b]{second half of the third observing run}
\acrodef{O3GK}[O3GK]{observing run}
\acrodef{O4}[O4]{fourth observing run}
\acrodef{O4a}[O4a]{first part of the fourth observing run}
\acrodef{O4b}[O4b]{second part of the fourth observing run}
\acrodef{O4c}[O4c]{third part of the fourth observing run}
\acrodef{IR1}[IR1]{intermediate run 1}
\acrodef{O5}[O5]{fifth observing run}

\acrodef{BH}[BH]{black hole}
\acrodef{BBH}[BBH]{binary \ac{BH}}
\acrodef{BNS}[BNS]{binary \ac{NS}}
\acrodef{IMBH}[IMBH]{intermediate-mass \ac{BH}}
\acrodef{NS}[NS]{neutron star}
\acrodef{BHNS}[BHNS]{\ac{BH}--\ac{NS} binary}
\acrodefplural{BHNS}[BHNSs]{\ac{BH}--\ac{NS} binaries}
\acrodef{NSBH}[NSBH]{\ac{NS}--\ac{BH} binary}
\acrodefplural{NSBH}[NSBHs]{\ac{NS}--\ac{BH} binaries}
\acrodef{PBH}[PBH]{primordial \ac{BH}}
\acrodef{CBC}[CBC]{compact binary coalescence}

\acrodef{GW}[GW]{gravitational wave\protect\protectedacused{GWH}}
\acrodef{GWH}[GW]{gravitational-wave\protect\protectedacused{GW}}
\acrodef{IFO}[IFO]{interferometer}
\acrodef{SNR}[SNR]{signal-to-noise ratio}
\acrodef{FAR}[FAR]{false-alarm rate}
\acrodef{IFAR}[IFAR]{inverse false-alarm rate}
\acrodef{FAP}[FAP]{false alarm probability}
\acrodef{PSD}[PSD]{power spectral density}
\acrodef{GR}[GR]{general relativity}
\acrodef{NR}[NR]{numerical relativity}
\acrodef{PN}[PN]{post-Newtonian}
\acrodef{EOB}[EOB]{effective-one-body}
\acrodef{ROM}[ROM]{reduced-order model}
\acrodef{IMR}[IMR]{inspiral--merger--ringdown}
\acrodef{PDF}[pdf]{probability density function}
\acrodef{PE}[PE]{parameter estimation}
\acrodef{CI}[CI]{credible interval}
\acrodef{CL}[CL]{credible level}
\acrodef{EOS}[EoS]{equation of state}
\acrodef{KLD}[KLD]{Kullback--Leibler divergence}
\acrodef{JSD}[JSD]{Jensen--Shannon divergence}
\acrodef{GCN}[GCN]{General Coordinates Network}
\acrodef{GWTC}[GWTC]{Gravitational-Wave Transient Catalog}
\acrodef{GWOSC}[GWOSC]{Gravitational Wave Open Science Center}
\acrodef{WDM}[WDM]{Wilson--Daubechies--Meyer}
\acrodef{DQR}[DQR]{data-quality report}

\acrodef{CWB}[cWB]{coherent WaveBurst}
\acrodef{LAL}[LAL]{\ac{LIGO} algorithm library}

\acrodef{CHRoCC}{central heating radius of curvature correction}
\acrodef{NonSENS}{non-stationary estimation and noise subtraction}
\acrodef{RF}{radio frequency}
\acrodef{PNC}{phase noise cancellation}
\acrodef{ASC}{alignment sensing and control}
\acrodef{WFS}{wave-front sensing}
\acrodef{BPC}{beam position control}
\acrodef{ADS}{alignment dither systems}
\acrodef{OMC}{output mode cleaner}
\acrodef{LVDTs}{linear variable differential transformers}
\acrodef{GAS}{geometrical anti-spring}

\acrodef{PTA}{Pulsar Timing Array}

\newcommand\gwtc[1][?]{\mbox{GWTC\if#1?\else-#1\fi}}
\newcommand\thisgwtcversionmajor{5}
\newcommand\thisgwtcversionminor{0}
\newcommand\thisgwtcversionfull{\thisgwtcversionmajor.\thisgwtcversionminor}
\newcommand\thisgwtcversion\thisgwtcversionfull

\DeclareSIUnit\parsec{pc}
\DeclareSIUnit\Mpc{\mega\parsec}
\DeclareSIUnit\yr{yr}
\DeclareSIUnit\GpcCubedYear{\giga\parsec\cubed\yr}

\newcommand{\OoneDuration}{{{\qty{129.66666666666666}{\day}}}}
\newcommand{\OoneDurationH}{{{\qty{27.696053240740742}{\day}}}}
\newcommand{\OoneDurationHFraction}{{{\qty{21.359424093116253}{\percent}}}}
\newcommand{\OoneDurationL}{{{\qty{16.804375}{\day}}}}
\newcommand{\OoneDurationLFraction}{{{\qty{12.959672236503856}{\percent}}}}
\newcommand{\OoneDurationHL}{{{\qty{49.004895833333336}{\day}}}}
\newcommand{\OoneDurationHLFraction}{{{\qty{37.79297879177378}{\percent}}}}
\newcommand{\OtwoDuration}{{{\qty{268.25001157407405}{\day}}}}
\newcommand{\OtwoDurationH}{{{\qty{37.82396990740741}{\day}}}}
\newcommand{\OtwoDurationHFraction}{{{\qty{14.100267763441556}{\percent}}}}
\newcommand{\OtwoDurationL}{{{\qty{33.490462962962965}{\day}}}}
\newcommand{\OtwoDurationLFraction}{{{\qty{12.48479460129118}{\percent}}}}
\newcommand{\OtwoDurationV}{{{\qty{1.668263888888889}{\day}}}}
\newcommand{\OtwoDurationVFraction}{{{\qty{0.6219063623146266}{\percent}}}}
\newcommand{\OtwoDurationHL}{{{\qty{103.01023148148148}{\day}}}}
\newcommand{\OtwoDurationHLFraction}{{{\qty{38.40083020948404}{\percent}}}}
\newcommand{\OtwoDurationHV}{{{\qty{1.7190277777777778}{\day}}}}
\newcommand{\OtwoDurationHVFraction}{{{\qty{0.6408304580084198}{\percent}}}}
\newcommand{\OtwoDurationLV}{{{\qty{2.2100115740740742}{\day}}}}
\newcommand{\OtwoDurationLVFraction}{{{\qty{0.8238626202123408}{\percent}}}}
\newcommand{\OtwoDurationHLV}{{{\qty{15.252222222222223}{\day}}}}
\newcommand{\OtwoDurationHLVFraction}{{{\qty{5.685823509465348}{\percent}}}}
\newcommand{\OthreeDuration}{{{\qty{361.0833333333333}{\day}}}}
\newcommand{\OthreeDurationH}{{{\qty{10.106608796296296}{\day}}}}
\newcommand{\OthreeDurationHFraction}{{{\qty{2.7989685103982356}{\percent}}}}
\newcommand{\OthreeDurationL}{{{\qty{9.700381944444445}{\day}}}}
\newcommand{\OthreeDurationLFraction}{{{\qty{2.686466266635895}{\percent}}}}
\newcommand{\OthreeDurationV}{{{\qty{25.040046296296296}{\day}}}}
\newcommand{\OthreeDurationVFraction}{{{\qty{6.934700105136293}{\percent}}}}
\newcommand{\OthreeDurationHL}{{{\qty{48.86767361111111}{\day}}}}
\newcommand{\OthreeDurationHLFraction}{{{\qty{13.533627586737442}{\percent}}}}
\newcommand{\OthreeDurationHV}{{{\qty{31.943993055555556}{\day}}}}
\newcommand{\OthreeDurationHVFraction}{{{\qty{8.846709362258636}{\percent}}}}
\newcommand{\OthreeDurationLV}{{{\qty{39.076203703703705}{\day}}}}
\newcommand{\OthreeDurationLVFraction}{{{\qty{10.821935020642613}{\percent}}}}
\newcommand{\OthreeDurationHLV}{{{\qty{154.28855324074075}{\day}}}}
\newcommand{\OthreeDurationHLVFraction}{{{\qty{42.729347770341306}{\percent}}}}
\newcommand{\OthreeGKDuration}{{{\qty{13.666666666666666}{\day}}}}
\newcommand{\OthreeGKDurationG}{{{\qty{4.501018518518518}{\day}}}}
\newcommand{\OthreeGKDurationGFraction}{{{\qty{32.93428184281843}{\percent}}}}
\newcommand{\OthreeGKDurationK}{{{\qty{0.8890046296296297}{\day}}}}
\newcommand{\OthreeGKDurationKFraction}{{{\qty{6.5049119241192415}{\percent}}}}
\newcommand{\OthreeGKDurationGK}{{{\qty{6.380150462962963}{\day}}}}
\newcommand{\OthreeGKDurationGKFraction}{{{\qty{46.68402777777778}{\percent}}}}
\newcommand{\OthreeaDuration}{{{\qty{183.0}{\day}}}}
\newcommand{\OthreeaDurationH}{{{\qty{5.58949074074074}{\day}}}}
\newcommand{\OthreeaDurationHFraction}{{{\qty{3.0543665249949403}{\percent}}}}
\newcommand{\OthreeaDurationL}{{{\qty{6.277905092592593}{\day}}}}
\newcommand{\OthreeaDurationLFraction}{{{\qty{3.430549230924914}{\percent}}}}
\newcommand{\OthreeaDurationV}{{{\qty{15.71050925925926}{\day}}}}
\newcommand{\OthreeaDurationVFraction}{{{\qty{8.584977737300141}{\percent}}}}
\newcommand{\OthreeaDurationHL}{{{\qty{25.9428125}{\day}}}}
\newcommand{\OthreeaDurationHLFraction}{{{\qty{14.176400273224044}{\percent}}}}
\newcommand{\OthreeaDurationHV}{{{\qty{17.467962962962964}{\day}}}}
\newcommand{\OthreeaDurationHVFraction}{{{\qty{9.545334952438777}{\percent}}}}
\newcommand{\OthreeaDurationLV}{{{\qty{25.315474537037037}{\day}}}}
\newcommand{\OthreeaDurationLVFraction}{{{\qty{13.833592643189638}{\percent}}}}
\newcommand{\OthreeaDurationHLV}{{{\qty{80.84550925925926}{\day}}}}
\newcommand{\OthreeaDurationHLVFraction}{{{\qty{44.17787391216353}{\percent}}}}
\newcommand{\OthreebDuration}{{{\qty{147.08333333333334}{\day}}}}
\newcommand{\OthreebDurationH}{{{\qty{4.517118055555556}{\day}}}}
\newcommand{\OthreebDurationHFraction}{{{\qty{3.071128423040604}{\percent}}}}
\newcommand{\OthreebDurationL}{{{\qty{3.422476851851852}{\day}}}}
\newcommand{\OthreebDurationLFraction}{{{\qty{2.326896443185395}{\percent}}}}
\newcommand{\OthreebDurationV}{{{\qty{9.329537037037037}{\day}}}}
\newcommand{\OthreebDurationVFraction}{{{\qty{6.343028013849543}{\percent}}}}
\newcommand{\OthreebDurationHL}{{{\qty{22.92486111111111}{\day}}}}
\newcommand{\OthreebDurationHLFraction}{{{\qty{15.586307837582625}{\percent}}}}
\newcommand{\OthreebDurationHV}{{{\qty{14.476030092592593}{\day}}}}
\newcommand{\OthreebDurationHVFraction}{{{\qty{9.842060119609695}{\percent}}}}
\newcommand{\OthreebDurationLV}{{{\qty{13.760729166666666}{\day}}}}
\newcommand{\OthreebDurationLVFraction}{{{\qty{9.355736543909348}{\percent}}}}
\newcommand{\OthreebDurationHLV}{{{\qty{73.44304398148148}{\day}}}}
\newcommand{\OthreebDurationHLVFraction}{{{\qty{49.93294774944916}{\percent}}}}
\newcommand{\OfouraDuration}{{{\qty{237.04166666666666}{\day}}}}
\newcommand{\OfouraDurationH}{{{\qty{33.34704861111111}{\day}}}}
\newcommand{\OfouraDurationHFraction}{{{\qty{14.06801136696549}{\percent}}}}
\newcommand{\OfouraDurationL}{{{\qty{36.99885416666667}{\day}}}}
\newcommand{\OfouraDurationLFraction}{{{\qty{15.60858674635261}{\percent}}}}
\newcommand{\OfouraDurationHL}{{{\qty{126.47008101851851}{\day}}}}
\newcommand{\OfouraDurationHLFraction}{{{\qty{53.35352336868421}{\percent}}}}
\newcommand{\OfourbDuration}{{{\qty{293.0833333333333}{\day}}}}
\newcommand{\OfourbDurationH}{{{\qty{7.846747685185186}{\day}}}}
\newcommand{\OfourbDurationHFraction}{{{\qty{2.6773094177487127}{\percent}}}}
\newcommand{\OfourbDurationL}{{{\qty{22.822141203703705}{\day}}}}
\newcommand{\OfourbDurationLFraction}{{{\qty{7.786911983066376}{\percent}}}}
\newcommand{\OfourbDurationV}{{{\qty{30.26615740740741}{\day}}}}
\newcommand{\OfourbDurationVFraction}{{{\qty{10.326809465137586}{\percent}}}}
\newcommand{\OfourbDurationHL}{{{\qty{21.566111111111113}{\day}}}}
\newcommand{\OfourbDurationHLFraction}{{{\qty{7.358354658326225}{\percent}}}}
\newcommand{\OfourbDurationHV}{{{\qty{22.05903935185185}{\day}}}}
\newcommand{\OfourbDurationHVFraction}{{{\qty{7.526541718004612}{\percent}}}}
\newcommand{\OfourbDurationLV}{{{\qty{64.21601851851852}{\day}}}}
\newcommand{\OfourbDurationLVFraction}{{{\qty{21.910498215019114}{\percent}}}}
\newcommand{\OfourbDurationHLV}{{{\qty{91.07658564814815}{\day}}}}
\newcommand{\OfourbDurationHLVFraction}{{{\qty{31.075320664707927}{\percent}}}}

\newcommand\PEpdfp{\ensuremath{p}}

\newcommand\PEpdf[2][?]{\ensuremath{\PEpdfp({#2}\ifx#1?\else | {#1}\fi)}}

\newcolumntype{L}{>{\centering\arraybackslash}m{3cm}} 

\newcommand{\GEO}{GEO\,600}

\newcommand{\samplingF}{\qty{16384}{\hertz}}
\newcommand{\samplingFs}{\qty{16}{\kilo\hertz}}
\newcommand{\subsamplingF}{\qty{4096}{\hertz}}
\newcommand{\subsamplingFs}{\qty{4}{\kilo\hertz}}

\newcommand{\mainPowerFUS}{\qty{60}{\hertz}}

\newcommand{\LIGOfileduration}{\qty{4096}{\second}}

\begin{document}

\title{Open Data from LIGO, Virgo, and KAGRA through the Second Part of the Fourth Observing Run}

\author[0000-0003-4786-2698]{A.~G.~Abac}
\affiliation{Max Planck Institute for Gravitational Physics (Albert Einstein Institute), D-14476 Potsdam, Germany}
\author{A.~Abe}
\affiliation{Department of Physics, Graduate School of Science, Osaka Metropolitan University, 3-3-138 Sugimoto-cho, Sumiyoshi-ku, Osaka City, Osaka 558-8585, Japan  }
\author{I.~Abouelfettouh}
\affiliation{LIGO Hanford Observatory, Richland, WA 99352, USA}
\author{F.~Acernese}
\affiliation{Dipartimento di Fisica ``E.R. Caianiello'', Universit\`a di Salerno, I-84084 Fisciano, Salerno, Italy}
\affiliation{INFN, Sezione di Napoli, I-80126 Napoli, Italy}
\author[0000-0002-8648-0767]{K.~Ackley}
\affiliation{University of Warwick, Coventry CV4 7AL, United Kingdom}
\author{A.~Adam}
\affiliation{OzGrav, University of Western Australia, Crawley, Western Australia 6009, Australia}
\author[0009-0004-2101-5428]{S.~Adhicary}
\affiliation{The Pennsylvania State University, University Park, PA 16802, USA}
\author{D.~Adhikari}
\affiliation{Max Planck Institute for Gravitational Physics (Albert Einstein Institute), D-30167 Hannover, Germany}
\affiliation{Leibniz Universit\"{a}t Hannover, D-30167 Hannover, Germany}
\author[0000-0002-5731-5076]{R.~X.~Adhikari}
\affiliation{LIGO Laboratory, California Institute of Technology, Pasadena, CA 91125, USA}
\author{V.~K.~Adkins}
\affiliation{Louisiana State University, Baton Rouge, LA 70803, USA}
% ...

\begin{abstract}
  LIGO, Virgo, KAGRA, and \GEO{} form a network of gravitational-wave observatories. Data and analysis 
  results from this network are made publicly available through the Gravitational Wave Open 
  Science Center (GWOSC). This paper describes open data from this network, including the addition of 
  data from the second part of the fourth observing run (O4b) and selected 
  periods from the preceding engineering run (ER16), which were collected from times spanning April 6th, 2024 
  to January 28th, 2025. The public data set includes calibrated strain time series for each instrument, 
  data from additional channels used for noise subtraction and detector characterization, and new analysis 
  data products in the online GWOSC release associated with version 5.0 of the Gravitational-Wave Transient Catalog. 
\end{abstract}

\section{Introduction}

The \acl{LIGO} \citep[\acs{LIGO};][]{2015CQGra..32g4001L}, Virgo \citep{2015CQGra..32b4001A}, 
KAGRA \citep{2021PTEP.2021eA101A} and \GEO{} \citep{2010JPhCS.228a2012L, 2014CQGra..31v4002A, 2016CQGra..33g5009D} are observing 
the gravitational-wave (GW) universe with unprecedented sensitivity \citep{GWTC:Introduction}. Data from 
these observatories are shared and jointly analyzed as described in a Memorandum of Agreement \citep{lvkmoa}, 
so that the instruments form a global network. The instruments operate in a series of observing runs, with 
breaks between observing periods for instrument upgrades and detector commissioning \citep{2020LRR....23....3A,2025PhRvD.111f2002C}. 
Following a proprietary period for data characterization and analysis by the LIGO--Virgo--KAGRA Collaboration (LVK),
data from each observing run are publicly released via the Gravitational Wave Open Science Center (GWOSC) for use 
by the entire scientific community.\footnote{GWOSC Home Page, \url{https://gwosc.org}} The schedule of data releases 
for LIGO is maintained in the LIGO Data Management Plan \citep{LIGO-DMP}.

Data from previous observing runs have been released with a history going back more than a decade, starting 
with data from initial LIGO released in 2014 \citep{Vallisneri:2014vxa}. Data from the advanced-detector era 
are organized into an ongoing series of observing runs, beginning with data from the first observing run (O1)
and \acl{O2} \citep[\acsu{O2};][]{2021SoftX..1300658A}, and continuing with data released from the 
\acl{O3} \citep[\acsu{O3};][]{2023ApJS..267...29A}.

The fourth observing run (O4) is divided into several parts, with the first part (O4a) spanning time from 
May 24, 2023 to January 16, 2024. The second part (O4b) spanned the time beginning on April 10, 2024 to 
January 28, 2025. This paper describes the publicly available data from the LVK with an emphasis on the 
new O4b public data release, which follows the structure of the previous O4a public data release \citep{2025arXiv250818079T}. As described 
in Subsection~\ref{sec:obstime}, this data release includes new data from both LIGO and Virgo for the second part (O4b).

Public data include time series strain data from the instruments, additional time series channels describing 
the instrument state, and segment information describing data quality of each instrument. In addition, the 
cumulative set of events identified by the LVK Collaboration, known as the Gravitational-Wave Transient Catalog (GWTC) is 
publicly available. The fifth version of the cumulative catalog \citep[GWTC-5.0;][]{GWTC:Introduction,GWTC:Methods, GWTC:Results} 
includes events from previous observing runs, those identified in O4b data, along with revised analysis results from O4a. 

Both O4a and O4b data releases include an electronic version of the catalogs that can be browsed online and 
queried through a Representational State Transfer Application Programming Interface (REST API). This paper is 
included in a suite of papers describing GWTC-5.0 \citep{GWTC:Introduction, GWTC:Methods, GWTC:Results} and is best understood in the 
context of the related papers.  

In Section~\ref{sec:dataoverview} we discuss the primary data products from GW observatories and their periods 
of data taking. In Section~\ref{sec:calib} we explain the basics of calibration of 
interferometer strain data. Section~\ref{sec:DQ} outlines some non-astrophysical noise sources that impact 
GW data analysis and noise mitigations. In Section~\ref{sec:data} we describe the details for the strain 
data available on GWOSC and alternative services. Section~\ref{sec:auxc} discusses additional instrumental 
monitoring channels used for noise subtraction and production of data-quality flags that are now also distributed 
by GWOSC.  Finally, the GWOSC Event Portal that provides electronic catalogs of detected GW transients
is described in Section~\ref{sec:portal}.

\section{Data Set Overview} \label{sec:dataoverview}

\subsection{Observing Time} \label{sec:obstime}

The instruments operated in a series of observing runs described in
\citet{GWTC:Introduction}, with the following
dates:
{
\addtolength{\leftmargini}{-0.5em}
\begin{itemize}
\item O1: 12 September 2015 0:00 UTC (GPS 1126051217)\newline
  to 19 January 2016 16:00 UTC (GPS 1137254417)
\item O2: 30 November 2016 16:00 UTC (GPS\ 1164556817)\newline
  to 25 August 2017 22:00 UTC (GPS 1187733618)
\item O3a: 1 April 2019 15:00 UTC (GPS 1238166018)\newline
  to 1 October 2019 15:00 UTC (GPS 1253977218)
\item O3b: 1 November 2019 15:00 UTC (GPS 1256655618)\newline
  to 27 March 2020 17:00 UTC (GPS 1269363618)  
\item O4a: 24 May 2023 15:00 UTC (GPS 1368975618)\newline
  to 16 January 2024 16:00 UTC (GPS 1389456018)
\item O4b: 10 April 2024 15:00 UTC (GPS 1396796418)\newline
  to 28 January 2025 17:00 UTC (GPS 1422118818)   
\end{itemize}
}

The O4a and O4b data releases also include additional data from engineering runs carried out immediately before these observing runs (see Appendix \ref{sec:O4a_bulk} and Section \ref{sec:O4b_bulk}).
Some observational data are also available from other times studied in 
targeted LVK publications,
including data from the \GEO{} \citep{2016CQGra..33g5009D} and KAGRA \citep{2021PTEP.2021eA101A} detectors,
as described in \citet{gwosc2017gw170817, gwosc2024geo600frbsb}.
In particular, GEO and KAGRA performed a joint observing run, called O3GK, from the 7 April 2020 8:00 UTC (GPS 1270281618) to 21 April 2020 0:00 UTC \citep[GPS 1271462418;][]{gwosc2022o3gk, 10.1093/ptep/ptac073}.

The observatories do not record astrophysical data at all times.
Many types of phenomena can interrupt data taking, including instrument
maintenance and detector improvements, seismic activity, power outages, and instrument
lock-loss for other reasons.
A comparison of the total amount of observing time for each run is
shown in Tables \ref{dutycycle} and \ref{dutycycle_O4}, where the engineering runs are not taken into account 
(for further details on this, see Sections \ref{sec:O4a_bulk} and \ref{sec:O4b_bulk}).  
When only one detector is listed in the tables, the observing time corresponds to periods in which that detector alone was taking data.
The
amount of calendar time for each observing run is shown at the top of
each table section (e.g., O4a spanned 237 calendar days).
O3 includes observing time from both O3a and O3b,
as well as a commissioning break with no observing time.

During O4a, Virgo was not observing due to
commissioning activities.
KAGRA and \GEO{} data were less sensitive than the
other observatories, and so were not used for most analyses.
Therefore, only data from LIGO are included in the O4a data release \citep{2025ApJ...995L..18A}.
The total coincident time with
both LIGO Hanford Observatory (LHO) and LIGO Livingston Observatory (LLO)
observing simultaneously during O4a was 126.5 days.
The O4b data release also includes Virgo data, provided from the offline analysis-ready reconstruction using the \texttt{V1:Hrec\_hoftRepro1A\_16384Hz} channel, alongside the LHO and LLO data.
The total coincident observing time for the three detectors amounts to \sisetup{round-mode=places,round-precision=1}\OfourbDurationHLV ays in addition to double detectors time (e.g., the period of time for which only LHO and LLO operated was of  \sisetup{round-mode=places,round-precision=1}\OfourbDurationHL ays).
The table shows similar statistics for each of O1, O2, and O3,
with additional information on past data releases available in
\citet{2021SoftX..1300658A} and \citet{2023ApJS..267...29A}.

\begin{deluxetable}{lrr}
  \sisetup{round-mode=places,round-precision=1,drop-zero-decimal=false}

  \tablecaption{
    The amount of observing time in days and the corresponding fraction of the
    total run time for each detector (IFO) combination from O1 to O3.
    The amount of calendar time for each observing run is shown at the
    top of each table section.
  }
  \label{dutycycle}

  \tablehead{\colhead{IFO Combination}&\colhead{Observing Time}&\colhead{Fraction}}

  \startdata
  \multicolumn{3}{c}{O1: {\OoneDuration}} \\
  LHO & {\OoneDurationH} & {\OoneDurationHFraction} \\
  LLO & {\OoneDurationL} & {\OoneDurationLFraction} \\
  LHO, LLO & {\OoneDurationHL} & {\OoneDurationHLFraction} \\
  \hline
  \multicolumn{3}{c}{O2: {\OtwoDuration}} \\
  LHO & {\OtwoDurationH} & {\OtwoDurationHFraction} \\
  LLO & {\OtwoDurationL} & {\OtwoDurationLFraction} \\
  Virgo & {\OtwoDurationV} & {\OtwoDurationVFraction} \\
  LHO, LLO & {\OtwoDurationHL} & {\OtwoDurationHLFraction} \\
  LHO, Virgo & {\OtwoDurationHV} & {\OtwoDurationHVFraction} \\
  LLO, Virgo & {\OtwoDurationLV} & {\OtwoDurationLVFraction} \\
  LHO, LLO, Virgo & {\OtwoDurationHLV} & {\OtwoDurationHLVFraction} \\
  \hline
  \multicolumn{3}{c}{O3a: {\OthreeaDuration}} \\
  LHO & {\OthreeaDurationH} & {\OthreeaDurationHFraction} \\
  LLO & {\OthreeaDurationL} & {\OthreeaDurationLFraction} \\
  Virgo & {\OthreeaDurationV} & {\OthreeaDurationVFraction} \\
  LHO, LLO & {\OthreeaDurationHL} & {\OthreeaDurationHLFraction} \\
  LHO, Virgo & {\OthreeaDurationHV} & {\OthreeaDurationHVFraction} \\
  LLO, Virgo & {\OthreeaDurationLV} & {\OthreeaDurationLVFraction} \\
  LHO, LLO, Virgo & {\OthreeaDurationHLV} & {\OthreeaDurationHLVFraction} \\
  \hline
  \multicolumn{3}{c}{O3b: {\OthreebDuration}} \\
  LHO & {\OthreebDurationH} & {\OthreebDurationHFraction} \\
  LLO & {\OthreebDurationL} & {\OthreebDurationLFraction} \\
  Virgo & {\OthreebDurationV} & {\OthreebDurationVFraction} \\
  LHO, LLO & {\OthreebDurationHL} & {\OthreebDurationHLFraction} \\
  LHO, Virgo & {\OthreebDurationHV} & {\OthreebDurationHVFraction} \\
  LLO, Virgo & {\OthreebDurationLV} & {\OthreebDurationLVFraction} \\
  LHO, LLO, Virgo & {\OthreebDurationHLV} & {\OthreebDurationHLVFraction} \\
  \hline
  \multicolumn{3}{c}{O3: {\OthreeDuration}} \\
  LHO & {\OthreeDurationH} & {\OthreeDurationHFraction} \\
  LLO & {\OthreeDurationL} & {\OthreeDurationLFraction} \\
  Virgo & {\OthreeDurationV} & {\OthreeDurationVFraction} \\
  LHO, LLO & {\OthreeDurationHL} & {\OthreeDurationHLFraction} \\
  LHO, Virgo & {\OthreeDurationHV} & {\OthreeDurationHVFraction} \\
  LLO, Virgo & {\OthreeDurationLV} & {\OthreeDurationLVFraction} \\
  LHO, LLO, Virgo & {\OthreeDurationHLV} & {\OthreeDurationHLVFraction} \\
  \hline
  \multicolumn{3}{c}{O3GK: {\OthreeGKDuration}} \\
  \GEO & {\OthreeGKDurationG} & {\OthreeGKDurationGFraction} \\
  KAGRA & {\OthreeGKDurationK} & {\OthreeGKDurationKFraction} \\
  \GEO, KAGRA & {\OthreeGKDurationGK} & {\OthreeGKDurationGKFraction} \\
  \enddata
\end{deluxetable}

\begin{deluxetable}{lrr}
  \sisetup{round-mode=places,round-precision=1,drop-zero-decimal=false}

  \tablecaption{
    The amount of observing time in days and the corresponding fraction of the
    total run time for each detector (IFO) combination in O4a and O4b.
    The amount of calendar time for each observing run is shown at the top of
    each table section.
  }
  \label{dutycycle_O4}

  \tablehead{\colhead{IFO Combination}&\colhead{Observing Time}&\colhead{Fraction}}

  \startdata
  \multicolumn{3}{c}{O4a: {\OfouraDuration}} \\
  LHO & {\OfouraDurationH} & {\OfouraDurationHFraction} \\
  LLO & {\OfouraDurationL} & {\OfouraDurationLFraction} \\
  LHO, LLO & {\OfouraDurationHL} & {\OfouraDurationHLFraction} \\
  \hline
  \multicolumn{3}{c}{O4b: {\OfourbDuration}} \\
  LHO & {\OfourbDurationH} & {\OfourbDurationHFraction} \\
  LLO & {\OfourbDurationL} & {\OfourbDurationLFraction} \\
  Virgo & {\OfourbDurationV} & {\OfourbDurationVFraction} \\
  LHO, LLO & {\OfourbDurationHL} & {\OfourbDurationHLFraction} \\
  LHO, Virgo & {\OfourbDurationHV} & {\OfourbDurationHVFraction} \\
  LLO, Virgo & {\OfourbDurationLV} & {\OfourbDurationLVFraction} \\
  LHO, LLO, Virgo & {\OfourbDurationHLV} & {\OfourbDurationHLVFraction} \\
  \enddata
\end{deluxetable}

\subsection{Time-series Data} \label{subsec:tsdata}

The primary data product from
a GW observatory is the calibrated strain data,
often called $h(t)$.  The strain data are recorded
as a time series, so that each time sample records one measurement
of the fractional difference in lengths of the interferometer arms,
$\Delta L / L$.
The LIGO, Virgo, and KAGRA observatories typically use a sampling rate of \samplingF{},
corresponding to a data rate of $\sim$4 TB per year per instrument
for the strain data. This is a small fraction of the total data rate of several PB per year,
which includes hundreds of thousands of diagnostic channels to monitor the state of the instrument
and the local environment.

\begin{figure*} 
\begin{center}
\includegraphics[width=\linewidth]{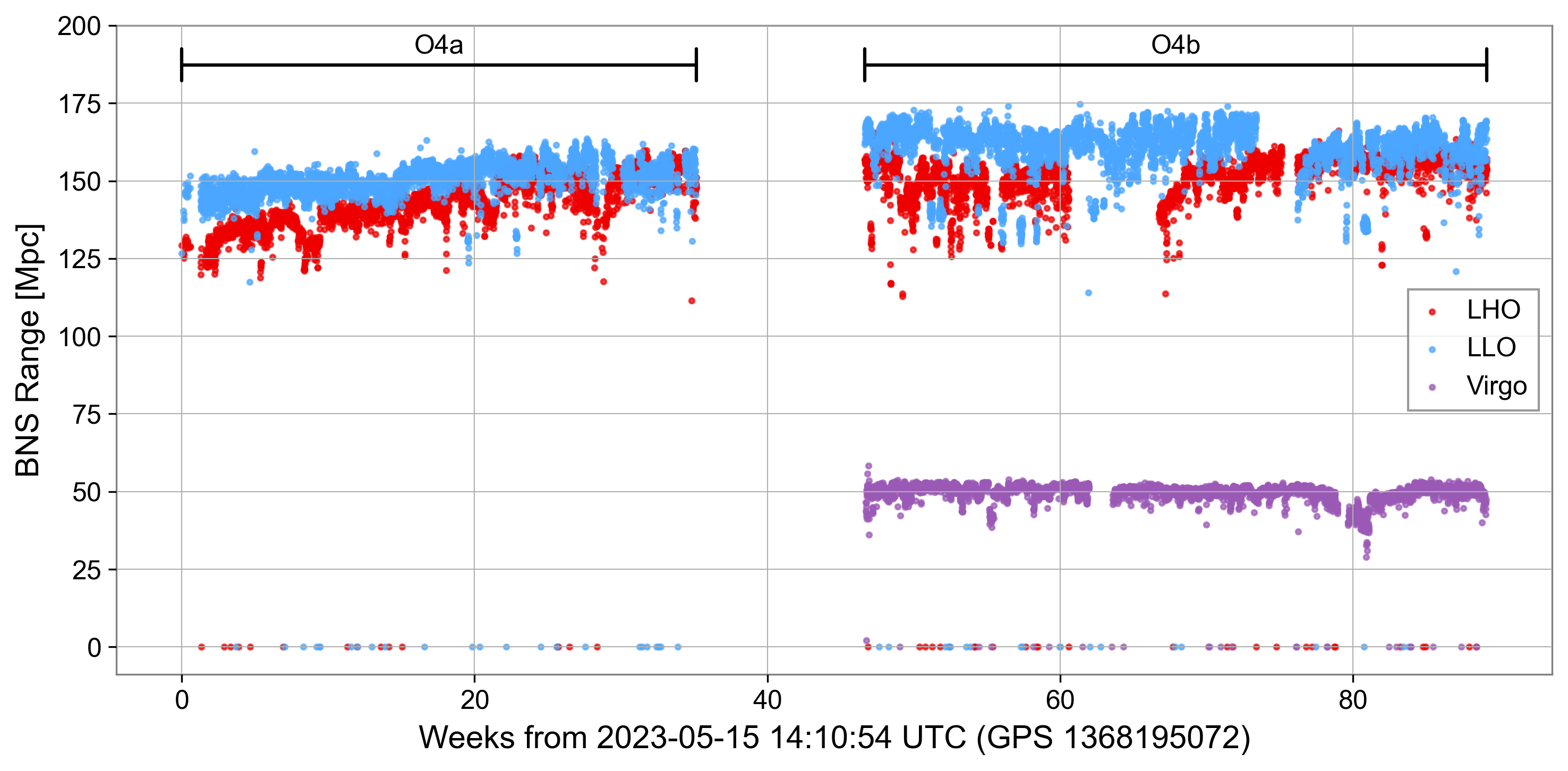}
\caption{
\label{sensemon}
BNS inspiral range over time for O4a and O4b.  Each point corresponds to a time
interval of \LIGOfileduration{}, corresponding to the time range of one strain
data file. If this time range contains a period of data too short to make a reliable inspiral range estimate, the range is set to zero.
The fast Fourier transform window length used is \qty{8}{\second} with a \qty{4}{\second} overlap.
These inspiral range values are stored as metadata for each strain data file
in the GWOSC database and are accessible via the GWOSC website.   
}
\end{center}
\end{figure*}

For current instruments, the strain time series consists primarily of instrumental noise at most times \citep{2020CQGra..37e5002A}.
The noise levels in the instruments fluctuate over time,
so that the sensitivities of the instruments vary.
One long-used figure of merit for the instrument sensitivity is the
binary neutron star (BNS) inspiral range \citep{1993PhRvD..47.2198F,2021CQGra..38e5010C}, which is briefly
summarized in \citet{GWTC:Introduction},
and gives an estimate of the distance to which a detector could detect
a BNS coalescence with a signal-to-noise ratio (SNR) of 8 after averaging over sky location and
orientation.
The variability in the BNS inspiral range over the course of O4a and O4b is
shown in Figure~\ref{sensemon}.
The plot shows that the LIGO observatories frequently operated near their peak sensitivity
of around \qty{160}{\Mpc} in O4a, with a slight improvement in O4b when Virgo joined with a sensitivity slightly above \qty{50}{\Mpc}. 
However, there are also times
when elevated noise levels cause the sensitivity to drop, as well as times
when the detectors are not operating at all or operating for periods too short to make a reliable inspiral range estimate, and in this case the
inspiral range is set to zero.
This variability in instrument sensitivity is an important
feature of the
strain data, and needs to be properly accounted for in any analysis.
Over the course of O1 through O4, the typical BNS inspiral range of the instruments
generally improved for each run
due to occasional hardware upgrades and commissioning work 
at the observatories
\citep{
2019PhRvL.123w1107T,
2020PhRvD.102f2003B,
2021ApOpt..60.4047B,2019PhRvD..99d2001D,
2021CQGra..38b5016S,
Soni:2023kqq,
2024Sci...385.1318J,
2025PhRvD.111f2002C}.
The evolution of this range is
described in \citet{2020LRR....23....3A} and \citet{GWTC:Introduction}.

In addition to the strain time series, the observatories also record
a large number of auxiliary channels to measure the state of each instrument
and the local environment \citep{2021CQGra..38n5001N,2024arXiv241204638H}.
These channels record quantities like
laser-power levels,
temperatures, seismic motion, angular alignments of mirrors, and many other
types of information that can be used to assess instrument
performance \citep{GraefRollins:2016xhy,2025CQGra..42h5016S}.
A subset of these auxiliary channels is included in the O3, 
O4a and O4b data releases, and is described in Section~\ref{auxrelease}.

\section{Calibration} \label{sec:calib}

Ground-based interferometers use feedback control loops to hold the differential arm degree of freedom, $\Delta L(t) = \Delta L_x(t) - \Delta L_y(t)$, in resonance.
Calibration refers to the reconstruction of the external differential arm motion $\Delta L(t)$ from the measured differential arm error signal $d_{\rm err}$ using a model of the true interferometer response function $R$~\citep{2017PhRvD..95f2003A,2018CQGra..35i5015V}.
In the time domain, the differential arm length can be reconstructed through a convolution of the differential arm error signal with the interferometer response function,
\begin{equation}
\Delta L(t) = R(t) \circledast d_{\rm err} (t),
\end{equation} 
where $\circledast$ denotes convolution,
implemented using digital finite impulse response filters for the response function \citep{2018CQGra..35i5015V}.
The strain is defined as the ratio of the differential arm motion to the unperturbed arm length $L$,
\begin{equation}
  h(t) = \frac{\Delta L(t)}{L} ,
\end{equation}
where $L \simeq \qty{3994.5}{\meter}$ for LIGO \citep{2015CQGra..32g4001L} and $L \simeq \qty{3000}{\meter}$ for Virgo.

\subsection{Calibration Details for LIGO}

\begin{figure*}[htbp]
  \centering
	\includegraphics[width=0.75\linewidth]{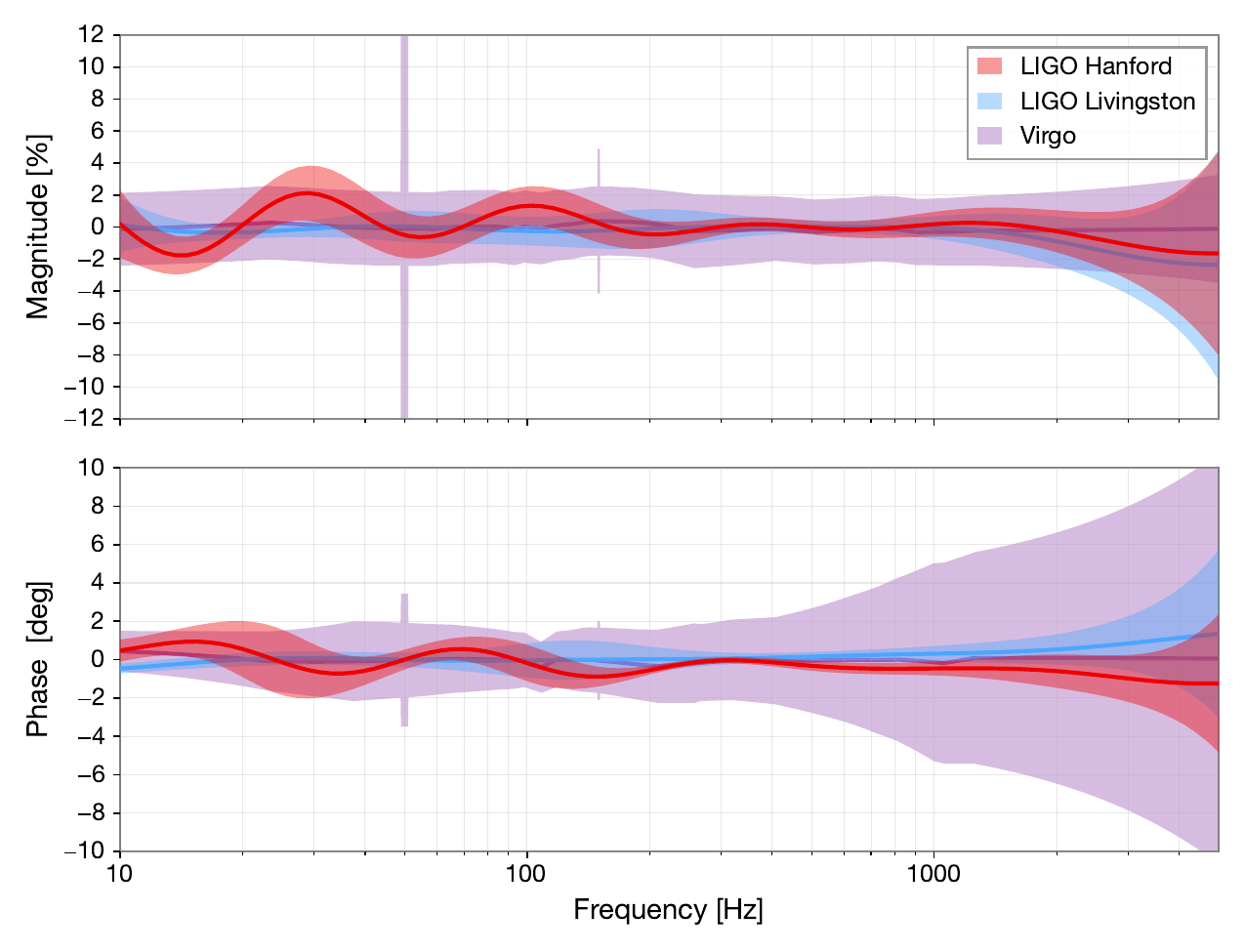}
    \caption{Estimated calibration errors and uncertainties at representative times in O4b (plotted as $R/R^{\rm (model)}-1$). 
  	Frequency-dependent calibration errors for LIGO Hanford (red) and Livingston (blue) are shown for a one-hour period starting at January 27, 2025 12:00:00 UTC. 
  	The Virgo frequency-dependent calibration uncertainty is shown for a one-month period starting at September 10, 2024 00:00:00 UTC. 
  	The magnitude of the calibration error is shown in the top panel, and the phase (in degrees) in the bottom panel. 
  	The solid curves denote the median systematic calibration error (called residual bias in section \ref{virgocalib}),
        and the shaded regions indicate the associated $1\sigma$ uncertainty. 
  	These uncertainties are representative of typical calibration performance for LIGO and Virgo during O4b.}
  \label{fig:calunc}
\end{figure*}

Imperfect calibration results in typically small errors in the measured strain.
Systematic errors in the imperfect, time-dependent and frequency-dependent calibration model $R^{\rm (model)}$ propagate to the reconstructed LIGO strain data through the ratio $R/R^{\rm (model)}$.
For LIGO in O4b, as in O4a, the broadband systematic errors and associated statistical uncertainties in the calibrated strain data are determined on an hourly cadence using a modified version of the methods described in \citet{2020CQGra..37v5008S}, incorporating additional continuous measurements of the systematic calibration error at discrete frequencies \citep{Wade:2025tgt}.
The best-estimated systematic errors in the reconstructed strain are left uncorrected in the LIGO strain data, but are reported in these hourly estimates.
The overall calibration uncertainties, including both systematic errors and associated statistical uncertainties, for all observing runs are available in the LIGO Document Control Center \citep{calerror, calerror-o4}.

Calibration methods for past observing runs are described in
\citet{2017PhRvD..95f2003A,2017PhRvD..96j2001C,2020CQGra..37v5008S,2021arXiv210700129S,2025arXiv250412657C} and \citet{2018CQGra..35t5004A,2022CQGra..39d5006A}.
During O4b, the majority of the final calibrated strain data for both LIGO observatories were computed in near real time (this calibration is designated as C00).
An exception was an approximately two-month period from September 3, 2024 19:03:10 UTC to November 14, 2024 19:20:14 UTC, during which the final strain data products were generated offline through a recalibration procedure (C01).

An overarching \texttt{ANALYSIS\_READY} strain data set, denoted \texttt{AR01}, was produced to provide a uniform data product for analysis. The \texttt{AR01} data set incorporates C01 strain data where available and otherwise uses the corresponding C00 strain data. This differs from O4a, where the \texttt{ANALYSIS\_READY} data set was denoted \texttt{C00\_AR}.

The recalibration was performed to address several issues identified in the online calibrated strain data. First, the subtraction of the \mainPowerFUS{} power mains line and its harmonics~\citep{Viets:2019} was not functioning as intended at both LIGO observatories throughout this period; consequently, this subtraction was disabled in the recalibration. Second, at LIGO Hanford, configuration changes to the signal-recycling cavity (SRC) around September 2024, were not fully captured in the online calibration model and are incorporated in the recalibrated data. Finally, a short miscalibration affecting LIGO Hanford on September 25, 2024, lasting approximately two hours, was corrected in the recalibration; this interval includes the time surrounding GW240925\_005809, for which astrophysical calibration provided an independent cross-check of the Hanford calibration~\citep{2026arXiv260511703T}.

The photon-calibrator system was used in O4b to inject continuous, sinusoidal excitations at eight frequencies throughout the run
\citep{2016RScI...87k4503K, Bhattacharjee_2024}.
The systematic error in the calibration at these frequencies was then inferred through the transfer function of the photon-calibrator excitation strain and the reconstructed strain \citep{Wade:2025tgt}.
The measured systematic error at each of these frequencies was included in the modeled systematic error.
Example uncertainty envelopes for the LHO and LLO calibrated strain data toward the end of O4b are shown in Figure~\ref{fig:calunc}.
The solid curves and shaded regions represent the median systematic calibration error and the associated $1\sigma$ uncertainty, respectively, over a one-hour period and are representative of typical calibration conditions during O4b.
The calibrated strain data for LIGO in O4b should be considered valid for
analyses within the provided hourly uncertainty envelopes in the range
\qtyrange{10}{5000}{\hertz}.

\subsection{Calibration Details for Virgo}
\label{virgocalib}

The Virgo $h(t)$ values are reconstructed by subtracting from the dark-fringe output signal the contribution of each mirror and marionette controls
involved in the differential arm length control loop (marionette is an upper part of the mirror suspension system
used for alignment controls and low frequency longitudinal controls).
The main contributions are the longitudinal controls of the north end, west end and beam splitter mirrors and,
below \qty{100}{\hertz}, the north input and west input marionette controls. 
Each contribution is obtained using actuator models
and optical response models adjusted online via permanent sinusoidal signals injected around \qty{70}{\hertz} and \qty{360}{\hertz}
via mirror actuators.
The actuator model is defined by the calibration measurements that take the photon calibrator \citep{PhdGrimaud,PCalO3}
and Newtonian calibrator \citep{NCalO3,NCalO4,PhdSyx} as reference. 
The resulting $h(t)$ signal is post-processed online with the following steps:
\begin{itemize}
  \item The bias on $h(t)$ is strongly mitigated by an unbiasing operation done using a bias measurement taken during
        the previous month. The result is what is called the residual bias. The stability of the bias and residual bias
        is checked with an online monitoring done with the permanent injected sinusoidal signals \citep{Amaldi2025_verkindt,Amaldi2025_grimaud}.
  \item All calibration lines injected continuously to monitor the $h(t)$ bias or to get information from the Newtonian calibrator are subtracted.
  \item A decorrelating linear subtraction, based on the method described in \citet{2019CQGra..36e5011D}, is used to mitigate
        any residual broadband noises below \qty{200}{\hertz} coming from the controls of the Michelson and signal recycling cavity degrees of freedom.
        During O4b, this noise subtraction allowed a gain of about \qty{0.5}{\Mpc} on the BNS inspiral range.
\end{itemize}

\begin{figure*}[htbp]
  \centering
 \includegraphics[width=\textwidth]{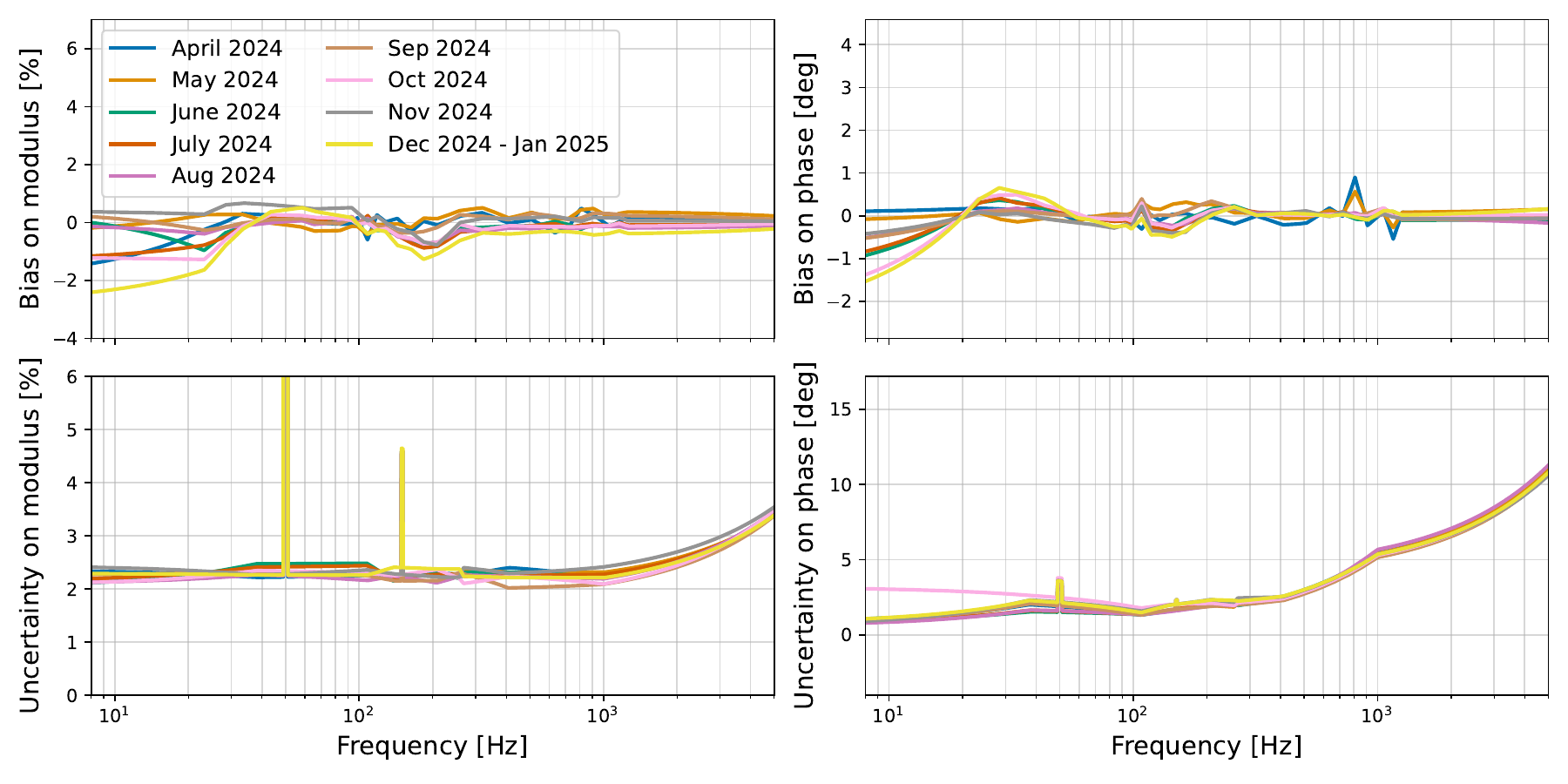}
  \caption{Virgo frequency-dependent calibration uncertainty for each month of O4b.
The magnitude and the phase uncertainties are stable from one month to an other.}
  \label{fig:virgostabunc}
\end{figure*}

An $h(t)$ offline production has also been done, within the \texttt{ANALYSIS\_READY} frames production.
It includes the updated state vector (summarized data quality information), the updated $h(t)$ uncertainties and residual bias on a monthly basis
and the reprocessing of $h(t)$ whenever needed (the first month of O4b); see Section \ref{sec:AlternateStrain} for further details on the \texttt{ANALYSIS\_READY} frames.
An example of the $h(t)$ uncertainties provided in the \texttt{ANALYSIS\_READY} frames is shown in purple in Figure~\ref{fig:calunc}.
The stability of these frequency-dependent uncertainties over O4b is illustrated by Figure~\ref{fig:virgostabunc}.

Details about the calibration procedure, the $h(t)$ online and offline
reconstruction and the estimation of $h(t)$ uncertainties and residual bias can
be found in \citet{O4Virgocal}.
Details about the Virgo \texttt{ANALYSIS\_READY} frames production can be found in \citet{Virgonote_ARframes}.

\section{Data Quality} \label{sec:DQ}

Noise in the strain data is sometimes approximated as stationary and Gaussian.
However, real strain data are
also affected by disturbances originating from instrumental and environmental sources \citep{2020CQGra..37e5002A, 2025CQGra..42h5016S}.
These artifacts can manifest as broadband, non-Gaussian, short-duration features
known as \textit{glitches} \citep{2018RSPTA.37670286N,2023CQGra..40f5004G}, 
or as narrowband spectral features referred to as \textit{spectral lines} \citep{2018PhRvD..97h2002C, O4_narrow_spectral_artifacts}.
In addition, GW observatories can experience periods of degraded performance due to
issues in control systems, calibration procedures, or environmental disturbances.
These episodes can compromise data quality over extended durations, reducing the sensitivity
of the detectors and potentially biasing astrophysical parameter estimation
\citep{2018PhRvD..98h4016P, 2022PhRvD.105j3021M,2022PhRvD.105b4066K, Ghonge:2023ksb, lecoeuche2026coalescing}.
To support robust analyses, the LVK employs a suite of data-quality
metrics designed to identify and allow mitigation of the impact
of instrumental and environmental noise.
An overview of LIGO detector characterization efforts is available in \citet{2025CQGra..42h5016S}.
Detector characterization methods for past observing runs are described in
\citet{2021CQGra..38m5014D,2023CQGra..40r5006A,2015CQGra..32x5005N,2016CQGra..33m4001A,2018CQGra..35f5010A, 2018PhRvD..97h2002C, 2021CQGra..38n5001N}; \citet{acernese2023virgo}; and \citet{2021PTEP.2021eA102A}.

Searches for different types of signals exhibit distinct susceptibility
to different noise
sources, but detector characterization is needed in all cases to remove
signals of terrestrial origin and strengthen the confidence of detections.
Data-quality products are employed across four broad categories of searches,
each with customized detector-characterization methods \citep{Caudill:2021afu}.
Compact binary coalescence (CBC) searches \citep{GWTC:Methods}
target GW signals from the coalescence of neutron stars
and/or black holes by applying matched-filtering techniques with modeled
waveform templates.
GW burst (BURST) analyses \citep{powell2025dawesreviewgravitationalwaveburst} aim to identify
short-duration transients without strong assumptions on
the signal morphology, e.g., by detecting excess power in the time--frequency
representation of the strain data.
Continuous wave (CW) searches \citep{Riles2023} focus on
long-duration, nearly monochromatic signals such as GWs emitted by non-axisymmetric
rotating neutron stars and various dark matter scenarios. Stochastic (STOCH) searches \citep{Romano2017,allen1997stochastic} aim to detect a diffuse
GW background resulting from the superposition of numerous unresolved sources or early-universe physics.

\subsection{Hardware Injections}

Simulated GW signals, known as \textit{hardware injections} \citep{2017PhRvD..95f2002B},
are introduced by physically displacing
the interferometer test masses to
mimic true astrophysical events and characterize the performance of the detectors and analyses.
Hardware injections with 
signals that mimic expected astrophysical sources are labeled by the simulation type, corresponding
to the different classes of potential sources (CBC, BURST, STOCH, or CW).
Detector characterization
\emph{safety injections} are labeled as DETCHAR injections.  These are used to test couplings
between channels used to monitor instrumental and environmental noise sources.

We identify auxiliary channels that can be used to indicate a detector problem, for example,
when they display excess noise levels.
However, some channels are \textit{unsafe} for identifying noise because they
respond to astrophysical signals and therefore a response in the
auxiliary channel could indicate a real signal.  
An auxiliary channel is said to be safe to use for identifying noise transients only if it does not respond to astrophysical signals.
Safety injections are used to identify unsafe channels. They are performed by actuating the end
test-mass mirror using a photon calibrator,
thereby generating a response in the primary strain channel \citep{2016RScI...87k4503K}.
Channels that exhibit a response to these injections are considered unsafe to use for vetoes,
and are identified following the method in \citet{2021PhRvD.103d2003E}.

Hardware injection times are available on the GWOSC website via a tool called 
Timelines{.}\footnote{Timelines App, \url{https://gwosc.org/timeline/}}
Additional notes about hardware injections are also provided in
the documentation for each data release.
In O4b there were no injections labeled
CBC, BURST, DETCHAR, or STOCH during observing mode.
The only hardware injections present in the released data are CW injections for the LIGO detectors, consisting of simulations of 18 spinning neutron stars whose parameters are provided on the GWOSC website.\footnote{O4b Injections, \url{https://gwosc.org/O4/o4_inj}}
Their GW frequencies span the range \qtyrange{12}{3000}{\hertz}.
No GW signal injections were performed for Virgo during O4b, apart from injections used for channel safety testing.
Injecting simulated CW signals allows, within the
uncertainties of the hardware injection system, a direct end-to-end assessment of degradation in
CW signal detection due to imperfect calibration or due to noise subtraction. These injections are almost always present (in O4b, only \qty{5}{\second} of the released LHO data and 6.23\% of LLO data do not contain CW injections), but because they are faint and nearly monochromatic, they have minimal impact on transient-signal searches.

\subsection{Data-quality Flags}

Data-quality flags are used to identify periods during which the strain data are affected by
data-quality issues.
These flags are used by analysis pipelines to exclude or down-weight compromised data segments.
Flags are classified in order of descending severity, such that the most severe problems are
labeled Category 1 (\texttt{CAT1}).
These flags mark intervals of time affected by well-understood and severe data-quality issues
(hardware faults, control-system failures, or known periods of malfunction)
during which the strain data are not considered suitable for astrophysical analyses; these times are systematically excluded from all
GW searches and parameter-estimation studies.
Some searches also apply the less severe Category 2 (\texttt{CAT2}),
 which are typically shorter in duration than \texttt{CAT1} \citep{GWTC:Methods}.
Category 3 (\texttt{CAT3}) flags are used to identify possible issues of unknown origin which are found through statistical studies. 
The \texttt{CAT2} flags were applied only to BURST searches.
The methodology for identifying and applying these flags is described in detail for LIGO in~\citet{2025CQGra..42h5016S}, while Virgo-specific details are provided in Section~\ref{sec:virgo_dq}.

For CBC analyses, an additional data-quality product used during O4 is the set of statistical flags derived from the \texttt{iDQ} supervised-learning framework \citep{Essick:2020qpo}. This pipeline evaluates the likelihood that transient noise is present in the strain data based on activity in auxiliary channels that are not expected to be sensitive to astrophysical sources.
The \texttt{iDQ} pipeline produces several statistical outputs sampled at \qty{128}{\hertz}.  The low-latency versions
of the \texttt{iDQ} outputs are available in the alternate strain release described in Section~\ref{sec:AlternateStrain};
the corresponding channel names are listed in Table~\ref{tab:idq_channels}.
The offline \texttt{iDQ} outputs are available in \citet{GWTC-5.0:DQ}. While the implementation of \texttt{iDQ} for the LIGO detectors remained unchanged in O4b relative to O4a, offline \texttt{iDQ} outputs were also produced for Virgo data during O4b. The \texttt{OK} flag indicates whether the \texttt{iDQ} outputs are reliable at a given time, taking the value 1 when valid and 0 otherwise. Other outputs include a normalized \texttt{RANK} score that reflects the raw classifier output, an estimated detection efficiency (\texttt{EFF}), the false alarm probability (\texttt{FAP}) of misclassifying clean times as glitchy, and the log likelihood-ratio (\texttt{LOGLIKE}), where positive values suggest the presence of a glitch and negative values suggest clean data. For CBC searches, the \texttt{iDQ} flags typically used are derived by thresholding the \texttt{LOGLIKE} channel; segments with $\texttt{LOGLIKE}>5$ are flagged as likely to be glitch-contaminated. Unlike \texttt{CAT1} and \texttt{CAT2} flags, \texttt{iDQ} flags are not applied universally. Search algorithms may choose to use them either as vetoes or to re-rank the significance of candidate events \citep{GWTC:Methods}.

\begin{table}[ht]
\centering
\caption{The \texttt{iDQ} channel names in the alternate strain
  release described in Section~\ref{sec:AlternateStrain},
  sampled at 128\,Hz. \textit{IFO} corresponds to H1 for LHO, L1 for
  LLO, or V1 for Virgo, and \texttt{AR} stands for \texttt{ANALYSIS\_READY}.
  }
\vspace{5pt}
\label{tab:idq_channels}
\begin{tabular}{ll}
\hline
			&	Channel name\\
\hline
\hline
OK Flag            &       \textit{IFO}\texttt{:IDQ-OK\_OVL\_10\_2048\_AR}\\
Rank               &       \textit{IFO}\texttt{:IDQ-RANK\_OVL\_10\_2048\_AR}\\
False Alarm Prob.  &       \textit{IFO}\texttt{:IDQ-FAP\_OVL\_10\_2048\_AR}\\
Efficiency         &       \textit{IfO}\texttt{:IDQ-EFF\_OVL\_10\_2048\_AR}\\
Log likelihood-ratio     &       \textit{IFO}\texttt{:IDQ-LOGLIKE\_OVL\_10\_2048\_AR}\\
\hline
\end{tabular}
\end{table}

\subsection{Data Quality Details for Virgo}\label{sec:virgo_dq}

For O4b, Virgo provides only \texttt{CAT1} data-quality flags. 
These are based on the automatic \texttt{CAT1} flags generated online during data taking, and used by online searches, complemented by a re-evaluation on the \texttt{ANALYSIS\_READY} data and by additional manual validation~\citep{2023CQGra..40r5006A}.
The manual checks refined the search for periods of excess noise.
Additionally, they targeted anomalous detector behavior or known instrumental issues that may have not been promptly flagged during online operations.
The resulting \texttt{CAT1} segments define the times considered unsuitable for scientific analysis in the Virgo open-data release.

Despite a significant improvement in the overall glitch rate with respect to O3, Virgo data during O4b are affected by a known family of recurrent transient noise events, commonly referred to as \emph{25-minute glitches}. 
These occur approximately every \qtyrange{23}{30}{\minute} and are characterized by relatively large amplitude, short duration, and a frequency content largely below \qty{200}{\hertz}~\citep{logbook:59763}. 
Due to their limited loudness and bandwidth, they are not flagged by the \texttt{CAT1} procedure.
However, their morphology and amplitude make them efficiently handled by the standard data-conditioning steps of transient-search pipelines, including automated data-quality checks and gating procedures, which remove or suppress their impact on searches~\citep{2020arXiv201015282G,2021PhRvD.104d2004C,2025CQGra..42j5009A}. 

\subsection{Spectral Line Catalog} 

Spectral lines are narrowband, often persistent features that appear in the amplitude spectral density of GW strain data. These features arise from a variety of instrumental and environmental sources, including mechanical resonances (e.g., violin-mode resonances from the suspension system), injected lines used for calibration, digital dither signals, and harmonics of the electrical mains. Although their impact on CBC and BURST searches is minimal, spectral lines are a significant concern for persistent searches.
Spectral artifacts that overlap with the target signal frequency can severely impact CW searches, which rely on coherence over long timescales. Similarly, STOCH searches are sensitive to lines, especially those of common origin or at harmonics of shared noise sources.

To support the identification and mitigation of these features,
the LVK maintains a curated catalog of instrumental spectral lines \citep{2018PhRvD..97h2002C, O4_lines_combs, O4_narrow_spectral_artifacts}.
The catalog includes both previously known artifacts and newly identified lines, and provides information such as line frequencies, harmonic structure, line widths, detector, temporal occurrence, and, where possible, a suspected source. This information is used by various search pipelines to identify contaminated frequency bins, apply vetoes, or model noise contamination. Spectral line catalogs are publicly available on the GWOSC website as part of the documentation of each run's data release.\footnote{GWOSC data releases, \url{https://gwosc.org/data}} 
The complete O4 spectral line list is documented in \citep{O4_lines_combs, O4_narrow_spectral_artifacts}. 

\subsection{Glitch Subtraction}\label{sub:glitch_subtraction}

Glitches represent a significant source of contamination in strain data, particularly affecting transient searches while having limited impact on persistent searches. In a small number of cases, a glitch occurs in close temporal and spectral proximity to a candidate GW signal, requiring targeted mitigation to reliably recover the astrophysical signal. Subtraction techniques are employed to remove localized excess power in the time--frequency domain \citep{2018PhRvD..98h4016P}.
One such method is \texttt{BayesWave} \citep{2021PhRvD.103d4006C,2022PhRvD.106d2006H},
which models and subtracts transient noise features from the data without requiring a specific model for the astrophysical signal. A complementary method is linear noise subtraction \citep{2019CQGra..36e5011D}, which removes known, repeatable noise contributions using information from auxiliary sensors that monitor the detector environment and instrumental systems.

During O4b, 23 candidate events with a false alarm rate (FAR) below 1 per year required such glitch mitigation
in at least one interferometer \citep{GWTC:Results}. These events are listed in Table~\ref{tab:glitch_sub}, along with their GPS time, affected interferometer(s), and the time--frequency region used to constrain the subtraction and assess residuals.
All glitch mitigated data listed in Table~\ref{tab:glitch_sub} were produced with the \texttt{BayesWave} pipeline.

\begin{table*}[t]
  \centering
  \caption{List of O4b events from \citet{GWTC:Results} with FAR < 1 per year for which glitch subtraction was performed using \texttt{BayesWave}  \citep{2021PhRvD.103d4006C}. For each event, we provide the GPS time, the interferometer (IFO) where subtraction was applied, and the time and frequency windows used for subtraction.  The time window indicates the approximate time of the glitch and is referenced from the indicated GPS time, with negative values before the merger and positive values after the merger. A similar table is included in \cite{GWTC:Results}.}
  \vspace{0.2cm}
  \begin{tabular}{l c c r r}
    \hline
    Event & GPS Time & IFO & Time (\unit{\second}) & Frequency (\unit{\hertz}) \\
    \hline\hline  
GW240514\_121713 & 1399724251.72 & LLO & $[0.70, 2.00]$ & $[20, 35]$ \\
GW240515\_005301 & 1399769599.11 & LLO & $[-4.01, -3.01]$ & $[10, 50]$ \\
GW240520\_213616 & 1400276194.97 & LLO & $[-0.28, -0.04]$ & $[60, 90]$ \\
GW240525\_031210 & 1400641948.64 & LLO & $[-4.64, -4.24]$ & $[10, 30]$ \\
GW240531\_075248 & 1401177186.46 & Virgo & $[-0.30, -0.20]$ & $[120, 200]$ \\
GW240601\_061200 & 1401257538.36 & LLO & $[0.52, 0.73]$ & $[200, 300]$ \\
GW240601\_231004 & 1401318622.03 & LLO & $[-5.30, -4.55]$ & $[20, 30]$ \\
GW240615\_160735 & 1402502873.33 & LLO & $[-0.05, -0.01]$ & $[100, 300]$ \\
\multirow{2}{*}{GW240621\_200935} &
\multirow{2}{*}{1403035793.53} &
LLO & $[0.30, 0.40]$ & $[200, 300]$ \\
&
& Virgo & $[-1.80, -1.60]$ & $[100, 200]$ \\
GW240629\_145256 & 1403707994.86 & LHO & $[-3.10, -2.90]$ & $[250, 400]$ \\
GW240705\_053215 & 1404192753.92 & Virgo & $[0.90, 1.10]$ & $[50, 100]$ \\
GW240902\_143306 & 1409322804.76 & Virgo & $[-0.10, 0.10]$ & $[300, 400]$ \\
GW240908\_082628 & 1409819206.41 & Virgo & $[-0.18, 0.03]$ & $[164, 210]$ \\
GW240908\_125134 & 1409835112.27 & LLO & $[0.60, 1.20]$ & $[20, 30]$ \\
GW240919\_061559 & 1410761777.19 & LHO & $[0.20, 0.80]$ & $[20, 30]$ \\
GW240920\_124024 & 1410871242.88 & LLO & $[0.002, 0.30]$ & $[20, 35]$ \\
GW240921\_201835 & 1410985133.79 & LLO & $[-4.60, -4.20]$ & $[10, 30]$ \\
GW240922\_142106 & 1411050084.39 & LHO & $[-0.40, -0.20]$ & $[150, 250]$ \\
GW240930\_035959 & 1411704017.96 & LLO & $[1.00, 1.80]$ & $[10, 800]$ \\
GW241009\_220455 & 1412546713.52 & LLO & $[-0.22, 0.58]$ & $[10, 30]$ \\
GW241102\_144729 & 1414594067.73 & LLO & $[-2.58, -1.63]$ & $[10, 30]$ \\
GW241109\_033317 & 1415158416.00 & LLO & $[0.60, 1.00]$ & $[10, 50]$ \\
GW241125\_010116 & 1416531694.78 & LHO & $[-0.60, -0.30]$ & $[20, 40]$ \\
    \hline
  \end{tabular}
  \label{tab:glitch_sub}
\end{table*}

\section{Strain Data} 
\label{sec:data}
Sections~\ref{sec:format} to~\ref{sec:previous_runs} describe the \emph{default strain-data release}, which
 can be downloaded directly from the GWOSC website
 or accessed via the \texttt{fetch\_open\_data} method of the \texttt{gwpy}
 package \citep{gwpy-software}.

Alternatively, especially for downloading large data sets, such as an entire observation run, users can download data from the network data server NDS2 \citep{nds2_wiki_page} or the Open Science Data Federation (OSDF) service that acts as a storage resource broker and optimizes data access through a worldwide network of data centers.\footnote{Large Scale Data and Computing Resources, \url{https://gwosc.org/osdf}} 
While much of the content in this section is similar across observing runs, Section~\ref{sec:previous_runs} describes differences between runs.

Data in the LVK archives are stored as gravitational-wave frame files (\texttt{gwf}; \citealp{frame}), a custom binary format developed within the GW community in which data are uniquely identified by a channel name and a frame type. 
Files for the default strain-data release are created by repackaging the original LVK data,
whose identifiers are listed in Table~\ref{tab:source} for all observing runs.
The best calibration version available is used in the default strain release.
In addition to the default strain-data release, several 
\emph{alternate strain} channels have also been released starting from O3 
(see Section~\ref{sec:AlternateStrain}).

\begin{table*}[t]
\centering
\caption{Channel names and frame types that allow tracing the provenance of strain data released on GWOSC. H1, L1, V1, G1 and K1 are short versions of the interferometer names. The attribute \texttt{CLEAN} in H1 and L1 indicates that noise subtraction \citep{2020PhRvD.101d2003V,2019CQGra..36e5011D,vietsdcc} was used. \texttt{AR} stands for  \texttt{ANALYSIS\_READY}, and indicates that these channels contain only data for times ready to be analysed. The other parts of the names refer in general to the calibration version.}
\label{tab:source}
\resizebox{\textwidth}{!}{\begin{tabular}{llll}
\hline
Run	&	IFO	&	Channel name					&	Frame type\\
\hline
\hline
O1      &       LHO      &       \texttt{H1:DCS-CALIB\_STRAIN\_C02}                      &       \texttt{H1\_HOFT\_C02}\\
O1      &       LLO      &       \texttt{L1:DCS-CALIB\_STRAIN\_C02}                      &       \texttt{L1\_HOFT\_C02}\\
\hline
O2      &       LHO      &       \texttt{H1:DCH-CLEAN\_STRAIN\_C02}              &       \texttt{H1\_CLEANED\_HOFT\_C02}\\
O2      &       LLO      &       \texttt{L1:DCH-CLEAN\_STRAIN\_C02}              &       \texttt{L1\_CLEANED\_HOFT\_C02}\\
O2      &       Virgo      &       \texttt{V1:Hrec\_hoft\_V1O2Repro2A\_16384Hz}    &       \texttt{V1O2Repro2A}\\
\hline
O3a	&	LHO	&	\texttt{H1:DCS-CALIB\_STRAIN\_CLEAN\_SUB60HZ\_C01}	&	\texttt{H1\_HOFT\_CLEAN\_SUB60HZ\_C01}\\
O3a	&	LLO	&	\texttt{L1:DCS-CALIB\_STRAIN\_CLEAN\_SUB60HZ\_C01}	&	\texttt{L1\_HOFT\_CLEAN\_SUB60HZ\_C01}\\
O3a	&	Virgo	&	\texttt{V1:Hrec\_hoft\_16384Hz}			&	\texttt{V1Online}\\
O3a(last two weeks)	&	Virgo	&	\texttt{V1:Hrec\_hoft\_V1O3ARepro1A\_16384Hz}	&	\texttt{V1O3Repro1A}\\
\hline
O3b	&	LHO	&	\texttt{H1:DCS-CALIB\_STRAIN\_CLEAN\_SUB60HZ\_C01}	&	\texttt{H1\_HOFT\_CLEAN\_SUB60HZ\_C01}\\
O3b	&	LLO	&	\texttt{L1:DCS-CALIB\_STRAIN\_CLEAN\_SUB60HZ\_C01}	&	\texttt{L1\_HOFT\_CLEAN\_SUB60HZ\_C01}\\
O3b	&	Virgo	&	\texttt{V1:Hrec\_hoft\_16384Hz}			&	\texttt{V1Online}\\
\hline
O3GK	&	\GEO	&	\texttt{G1:DER\_DATA\_HD\_CLEAN}		&	\texttt{G1\_RDS\_C02\_L3}\\
O3GK	&	KAGRA	&	\texttt{K1:DAC-STRAIN\_C20}			&	\texttt{K1\_HOFT\_C20}\\
\hline
O4a      &       LHO      &       \texttt{H1:GDS-CALIB\_STRAIN\_CLEAN\_AR}                      &       \texttt{H1\_HOFT\_C00\_AR}\\
O4a      &       LLO      &       \texttt{L1:GDS-CALIB\_STRAIN\_CLEAN\_AR}                      &       \texttt{L1\_HOFT\_C00\_AR}\\
\hline
O4b      &       LHO      &       \texttt{H1:DCS-CALIB\_STRAIN\_CLEAN\_AR01}                      &       \texttt{H1\_HOFT\_AR01}\\
O4b      &       LLO      &       \texttt{L1:DCS-CALIB\_STRAIN\_CLEAN\_AR01}                      &        \texttt{L1\_HOFT\_AR01}\\
O4b      &       Virgo    &        \texttt{V1:Hrec\_hoftRepro1AR\_16384Hz}				&	    \texttt{HoftAR1U02}\\  
\hline
\end{tabular}
}
\end{table*}  

The released calibrated strain data of the observing runs are divided into files containing \LIGOfileduration{} of data each. 
When data are missing or do not meet the necessary quality standards for analysis,
strain values are recorded as \texttt{NaN}s.

The strain data are available at both the original sampling rate of \samplingF{}
and a reduced rate of \subsamplingF{}, referred as \samplingFs{} and \subsamplingFs{} in the remainder of this paper. 
The downsampling process is carried out using the standard decimation method
\texttt{scipy.signal.decimate}, from the Python library \texttt{SciPy}
\citep{2020NatMe..17..261V}. Before downsampling, an order 8 Chebyshev type I
filter \citep{Smith} is used as an anti-aliasing filter.

The maximum resolvable frequency in a given data set is constrained by the Nyquist–Shannon sampling theorem \citep{Nyquist1} and corresponds to half of the sampling rate. 
Due to the roll-off effect of the anti-aliasing filters applied during resampling, the valid frequency range is slightly lower than the Nyquist frequency. Consequently, for data sampled at \subsamplingFs{}, frequencies above approximately \qty{1700}{\hertz} are increasingly attenuated by the anti-aliasing filter, which can suppress but not entirely eliminate usable signal content. 
For data sampled at \samplingFs{}, however, the limiting factor is given by the calibration which restricts the valid data to \qty{5}{\kilo\hertz}, as described in Section~\ref{sec:calib}.

These considerations on the maximum frequency are a crucial factor to consider when selecting a data set to download. 
On the other hand, data sampled at higher rates require increased storage space and longer download times. Users should, therefore, select the data set that best aligns with their specific requirements.

\subsection{Structure of the Default Strain-Data Files}
\label{sec:format}
The default strain files include not only the strain data but also information on data quality and injections. 
The main file formats used for the GWOSC open data are the Hierarchical Data Format (\texttt{hdf}; \citealp{hdf5}), a data format designed for storing and organizing large data sets that is easily readable by many programming languages, and the \texttt{gwf} format, which is the file
format used in the LVK archives as already mentioned.

For all available file formats, file names in the default strain release contain the following information:
\begin{itemize}
\item \textit{Obs}: the observatory, i.e. the site, that is indicated by one letter and can have values L for LIGO Livingston, H for LIGO Hanford, V for Virgo, G for \GEO{} or K for KAGRA.
\item \textit{IFO}: the interferometer label, created by adding a number to the letter that indicates the observatory. This choice allows the identification of multiple detectors installed in the same site, as was the case for Initial LIGO \citep{2009RPPh...72g6901A}. For the LVK observation runs it can have values H1, L1, V1, G1 or K1.
\item \textit{Run}: the observing run name, e.g., O4b. 
\item \textit{s}\texttt{KHZ} or \textit{s}: the sampling rate in \unit{\kilo\hertz}, \textit{s} can have values 4 or 16 (\subsamplingF{} or \samplingF{}).
\item \texttt{R}\textit{n} or \texttt{V}\textit{n}: release number or version number of the file. The letter \texttt{V} was used only for O1, then it was avoided because it could be confused with the Virgo interferometer name. 
\item \textit{GPSstart}: the starting time of the data contained in the file, as a 10-digit GPS value (in seconds).
\item \textit{Dur}: the duration in seconds of the file, typically \LIGOfileduration{}.
\item \textit{Ext}: the extension corresponding to the file format (\texttt{gwf}, \texttt{hdf} or \texttt{txt}).
\end{itemize}

The structure of the file names is:
\begin{center}
\textit{Obs}---\textit{FrameType}---\textit{GPSstart}---\textit{Dur}.\textit{Ext}
\end{center}
where the \textit{FrameType} is \textit{IFO}\_\texttt{GWOSC}\_\textit{Run}\_\textit{s}\texttt{KHZ}\_\texttt{R}\textit{n}.
 
The \texttt{hdf} files contain three folders (or groups).%
\begin{itemize}
\item The folder \emph{meta} hosts the metadata of the file: \emph{Description}, \emph{DescriptionURL}, \emph{Detector} (e.g., L1), \emph{Observatory} (e.g., L), \emph{Duration}, \emph{GPSstart}, \emph{UTCstart} (duration and starting time, using GPS and UTC standards, respectively, of the segment of data contained in a file), \emph{StrainChannel} and \emph{FrameType} used in the LVK archives as listed in Table~\ref{tab:source}.
	
\item The folder \emph{strain} contains the array \emph{Strain} of $h(t)$ values with useful attributes such as \emph{Xstart} and \emph{Xspacing} that define the GPS start time of the data contained in the array and the temporal distance between points in the array.

\item The folder \emph{quality} contains two subfolders: one for data quality and another for injections. Each subfolder includes a bitmask that indicates, for each second, the status of data quality or injections, along with a description of each bit in the mask (see Section \ref{sec:DqInj} for details).
\end{itemize}

The GWOSC \texttt{gwf} files include three channels (for strain data, data quality, and injections), whose names are listed in Table~\ref{tab:channels}.

\begin{table}[ht]
\centering
\caption{Channel names in the GWOSC \texttt{gwf} files. The nomenclature is similar to that described for the file name conventions: \textit{IFO} is the interferometer, $s$ is the sampling rate, $n$ is the version or release (until now only 1 has been used).}
\vspace{10pt}
\label{tab:channels}
\begin{tabular}{ll}
\hline
			&	Channel name\\
\hline
\hline
Strain                  &       \textit{IFO}\texttt{:GWOSC-}$s$\texttt{KHZ\_R$n$\_STRAIN}\\
Data-quality mask       &       \textit{IFO}\texttt{:GWOSC-}$s$\texttt{KHZ\_R$n$\_DQMASK}\\
Injections mask         &       \textit{IFO}\texttt{:GWOSC-}\texttt{$s$KHZ\_R$n$\_INJMASK}\\
\hline
\end{tabular}
\end{table}

Earlier releases used slightly different conventions described in Section~\ref{sec:previous_runs}.

\subsection{Data Quality and Injections in GWOSC Files}
\label{sec:DqInj}
The data-quality information is encoded in the default strain release
as a bitmask following the structure described in Table~\ref{tab:dq}, with a
\qty{1}{\hertz} sampling rate.\footnote{See tutorial 3 of \url{https://github.com/gwosc-tutorial/introduction_gwosc_data} to learn how to get data-quality information from GWOSC files.} 
The meaning of the data-quality flags is explained in Section~\ref{sec:DQ}. 
For template-based binary coalescence searches and minimally modeled searches, labeled respectively as CBC and BURST in Table~\ref{tab:dq}, the bitmask allows 3 categories of data quality (labeled with \texttt{CAT1}, \texttt{CAT2} and \texttt{CAT3}). 
We keep this structure even for runs in which not all the categories have been used. This choice avoids confusion that could arise if the same bit would have different meaning in different runs.
A bit value of 1 indicates the data has passed both the indicated level and
lower-level checks; for example, a 1 in the \texttt{CBC\_CAT2} bit indicates
the data pass both \texttt{CBC\_CAT1} and \texttt{CBC\_CAT2}.

The data-quality bit labeled as \texttt{DATA} is obtained requiring that both \texttt{CBC\_CAT1} and
\texttt{BURST\_CAT1} are satisfied. This is the criterion that defines what data are released in the \emph{default strain-data release}.

Starting from O4a, two additional bits were added to accommodate Category 1 data quality for STOCH
searches (\texttt{STOCH\_CAT1}) and for CW searches  (\texttt{CW\_CAT1}).	
Moreover, the Categories 2 and 3 have not been used for the CBC
data-quality flag so the corresponding segments coincide with those
marked as \texttt{CBC\_CAT1} (and also \texttt{DATA} in this case).
BURST searches from O4a did not apply any Category 3 flags. 
The CW data-quality flag \texttt{CW\_CAT1} is identical to
\texttt{CBC\_CAT1}.

\begin{table}[ht]
\centering
\caption{Meaning of the bits in the data-quality bitmask of the GWOSC files. Each bit has value 1 if the data quality passes the corresponding data quality requirement, otherwise the value is zero. Bits 7 and 8 were introduced in O4 and are not present in previous runs \citep{2021SoftX..1300658A, 2023ApJS..267...29A}.}
\vspace{10pt}
\label{tab:dq}
\begin{tabular}{lll}
\hline
Bit	&	Short name	&	Description	\\
\hline
\hline
0	& 	\texttt{DATA}		&	Data present	\\
1 	&	\texttt{CBC\_CAT1} 	&	Pass \texttt{CAT1} test for CBC search\\
2 	&	\texttt{CBC\_CAT2} 	&	\vtop{\hbox{\strut Pass \texttt{CAT1} and \texttt{CAT2} test}\hbox{\strut for CBC search}}\\
3 	&	\texttt{CBC\_CAT3} 	&	\vtop{\hbox{\strut Pass \texttt{CAT1} and \texttt{CAT2} and  \texttt{CAT3} test}\hbox{\strut for CBC search}} \\
4 	&	\texttt{BURST\_CAT1} 	&	Pass \texttt{CAT1} test for BURST search\\
5 	&	\texttt{BURST\_CAT2} 	&	\vtop{\hbox{\strut Pass \texttt{CAT1} and \texttt{CAT2} test}\hbox{\strut for BURST search}}\\
6 	&	\texttt{BURST\_CAT3} 	&	\vtop{\hbox{\strut Pass \texttt{CAT1} and \texttt{CAT2} and  \texttt{CAT3} test}\hbox{\strut for BURST search}}\\
7 	&	\texttt{STOCH\_CAT1} 	&	Pass \texttt{CAT1} test for STOCH search\\
8 	&	\texttt{CW\_CAT1} 	&	Pass \texttt{CAT1} test for CW search\\
\hline
\end{tabular}
\end{table}  

As discussed in Section~\ref{sec:DQ}, a crucial way to test the response of the detector is the use of \textit{hardware injections}.
In GWOSC files the information about injections is provided as \qty{1}{\hertz} time series containing a bitmask detailed in Table~\ref{tab:inj}.\footnote{See tutorials in \url{https://gwosc.org/tutorials/} to learn how to get hardware-injection information from GWOSC files.}
Five bits are used to distinguish injections relevant for astrophysical searches or detector-characterization studies.

\begin{table}[ht]
\centering
\caption{Meaning of the bits in the injections bitmask of the GWOSC files. The mask indicates when injections are not present, so the bit value is set to 1 to indicate no injection and 0 when there is an injection. \texttt{HW\_INJ} stands for hardware injection.}
\vspace{10pt}
\label{tab:inj}
\begin{tabular}{lll}
\hline
Bit	&	Short name				&	Description	\\
\hline
\hline
0	& 	\texttt{NO\_CBC\_HW\_INJ}		&	No CBC injections\\
1 	&	\texttt{NO\_BURST\_HW\_INJ} 		&	No BURST injections\\
2 	&	\texttt{NO\_DETCHAR\_HW\_INJ} 	&	No DETCHAR injections \\
3 	&	\texttt{NO\_CW\_HW\_INJ}  		&	No CW injections\\
4 	&	\texttt{NO\_STOCH\_HW\_INJ} 	&	No STOCH injections\\

\hline
\end{tabular}
\end{table}  

\subsection{Additional Segments in the O4b Strain-Data Release}
\label{sec:O4b_bulk}
As O4a (see Appendix \ref{sec:O4a_bulk}), also the O4b data release includes data from the engineering run just prior to the start
of the run that were used in the search for GW emitted from a supernova, in this
case the supernova SN 2024ggi~\citep{2025A&A...699A..60E}.
For this reason, the starting time of the data release is 6 April 2024 at 05:37:12 UTC (GPS 1396417050).
In addition, for the analysis of GW240925\_005809, \qty{10}{\second} of LHO data between 
25 September 2024 at 00:57:39 UTC (GPS 1411261077) and 25 September 2024 at 00:57:49 UTC (GPS 1411261087), 
initially labeled as out-of-observing mode, were, after further investigations,
promoted as analysis ready~\citep{2026arXiv260511703T}.

In this way, the amount of time added to the nominal O4b observing run is of \qty{2.9}{\day} for LHO, \qty{3.1}{\day} for LLO and  \qty{2.3} {\day} for Virgo and the total observing time for each of the three detectors is \qty{145.5}{\day}, \qty{202.8}{\day} and \qty{210.0}{\day}, respectively. Using these numbers as reference, we can calculate the percentage of observing time for which data quality fail the categories described in Table~\ref{tab:dq} and the result is in Table \ref{tab:dqO4b}.
\begin{table}[ht]
\centering
\caption{Percentage of the observing time in O4b for which data quality fail the categories described in Table~\ref{tab:dq}. 
For this calculation, the total amount of observing time is used as reference, including the engineering run just prior to the start of O4b, as discussed in the text.}
\vspace{10pt}
\label{tab:dqO4b}
\begin{tabular*}{\columnwidth}{@{\extracolsep{\fill}}llll}
\hline
Short name			&	 LHO				&	 LLO	&	 Virgo\\
\hline
\hline
\texttt{DATA}			&	0.7584\% 		&	0.7889\%	&	0.1274\% \\
\texttt{CBC\_CAT1} 		&	0.7584\%		&	0.7889\%	&	0.1274\%\\
\texttt{CBC\_CAT2} 		&	0.7584\%		&	0.7889\%	&	0.1274\%\\
\texttt{CBC\_CAT3} 		&	0.7584\%		&	0.7889\%	&	0.1274\%\\
\texttt{BURST\_CAT1} 	&	0.7584\%		&	0.7889\%	&	0.1274\%\\
\texttt{BURST\_CAT2} 	&	0.8755\%		 &	0.8363\%	&	0.1274\%\\
\texttt{BURST\_CAT3} 	&	0.8755\% 		&	0.8363\%	& 	0.1274\% \\
\texttt{STOCH\_CAT1} 	&	0.7584\% 		&	0.7889\%	&	0.1274\%\\
\texttt{CW\_CAT1} 		&	0.8105\% 		&	0.7889\%	&	0.1274\%\\
\hline
\end{tabular*}
\end{table}

\subsection{Data Releases from Previous Runs}
\label{sec:previous_runs}
The structure described here for the default strain data is largely valid for all 
data released to date \citep{2021SoftX..1300658A, 2023ApJS..267...29A, 2025arXiv250818079T}, with few differences.
These concern details in the naming conventions of the files and the channels within them. 
For example, the \textit{FrameType} mentioned in Section \ref{sec:format} is \textit{IFO}\_\texttt{LOSC}\_\textit{s}\_\texttt{V}\textit{n} 
in O1, the \emph{StrainChannel} and \emph{FrameType} have been present in files since O2, while the channel names 
in the \texttt{gwf} files used only in O1 are shown in Table~\ref{tab:channelsO1}.

The main difference that the user will notice is that up to O3, for each published GW detection, data snippets centered on the event detection time are also released via the GWOSC Event Portal, described in Section \ref{sec:portal}. 
These data snippets are provided as plain-text files in addition to the \texttt{hdf} and \texttt{gwf} files described above. 
The text files contain strain values in a single column.
As of O4, data snippets for most events are no longer published, but instead the associated
files from the default strain-data release are linked for each event in the Event Portal, if available at
the time of posting the event.
 
\begin{table}[ht]
\centering
\caption{Channel names in the GWOSC \texttt{gwf} files, used only in O1 for the \subsamplingFs{} release (\textit{IFO} is the interferometer name).}
\vspace{10pt}
\label{tab:channelsO1}
\begin{tabular}{ll}
\hline
			&	Channel name\\
\hline
\hline
Strain                  &       \textit{IFO}\texttt{:LOSC-}\texttt{STRAIN}\\
Data-quality mask       &       \textit{IFO}\texttt{:LOSC-}\texttt{DQMASK}\\
Injections mask         &       \textit{IFO}\texttt{:LOSC-}\texttt{INJMASK}\\
\hline
\end{tabular}
\end{table}

\subsection{Alternate Strain Release}
\label{sec:AlternateStrain}

The default strain files described in
Sections~\ref{sec:format}--\ref{sec:previous_runs} are prepared after each observing run
in order to provide a user-friendly file format that includes the preferred strain channel
and the final data quality and injection segments.
For O3 and O4, in addition to these user-friendly files,
the LVK Collaboration also releases several versions of the strain channel that more closely
matches the data available to LVK members before the public data release.
These data sets are called alternate strain releases and the frames contained in
them are called \texttt{ANALYSIS\_READY} frames.
Detailed documentation is available on the GWOSC website 
\citep{gwosc2021o3a,  gwosc2021o3b, gwosc2025o4a, gwosc2026o4b}.

Starting in O3, alternate strain releases are available for all times
the detectors passed the conditions for \texttt{ANALYSIS\_READY},
meaning that the detectors were nominally in a good working state.
This set of times includes a small amount of time that fails
\texttt{CAT1} data quality, and so there are some times in the alternate
strain releases that are not available in the default strain
files.
Files and strain channels are marked with the tag \texttt{AR},
to indicate that only \texttt{ANALYSIS\_READY} times are included.

For O3, those data sets were prepared after the run had ended, with
three channels of varying noise subtraction levels for LIGO,
and two channels for Virgo: a main and a short-duration one
covering a period when additional noise subtraction was needed.
See \citet{2023ApJS..267...29A} for details about those data sets.

Starting in O4a, the \texttt{ANALYSIS\_READY} frames are produced during the run,
with three channels of varying noise subtraction levels for LIGO.
See \citet{2025arXiv250818079T} for details about those data sets.
For O4b, analysis ready frames have been produced both for LIGO and Virgo.

For LIGO the process is the following.
The calibration pipeline described in Section~\ref{sec:calib} runs at all times
and writes out strain data
in \qty{1}{\second} long frame files in order to produce low-latency strain
data, including the preferred channel \texttt{GDS-CALIB\_STRAIN\_CLEAN}.
Low-latency analyses, including pipelines that produce public transient alerts,
use these low-latency frame files.\footnote{LIGO/Virgo/KAGRA Public Alerts User Guide, \url{https://emfollow.docs.ligo.org/userguide/}}
To produce the \texttt{ANALYSIS\_READY} frames, a frame aggregator
collects the \qty{1}{\second} files into \texttt{ANALYSIS\_READY} frame
files of up to \LIGOfileduration{} long, excludes any times that are not
marked as ready for analysis, and adds the tag \texttt{AR} to the
channel names.
The \texttt{ANALYSIS\_READY} frame files are then used for most offline LVK analyses,
so that the released product exactly matches data used inside the collaboration.
Those \texttt{ANALYSIS\_READY} frames are called \texttt{C00\_AR01}.

As explained in Section~\ref{sec:calib}, another version of the
\texttt{ANALYSIS\_READY} frames
was produced by incorporating the recalibrated data \texttt{C01}.
Those \texttt{ANALYSIS\_READY} frames are called \texttt{AR01}.

\begin{table*}[ht]
  \centering
  \caption{Alternate strain channels in the O4b release and the layers of noise subtraction applied to them.
    \texttt{NOLINES} means that narrowband noise features have been subtracted, while
    \texttt{CLEAN} means that broadband noise features have also been subtracted and
    \texttt{GATED} means the removal of loud glitches. For the first two sets of channels, \textit{IFO} can be H1 or L1.
  }
  \label{tab:o4archannels}
  \begin{tabular}{llccc}\hline
    Frame & Channel name & Narrowband        & Broadband         & Glitch  \\
          &              & noise subtraction & noise subtraction & removal \\ \hline \hline
    \multirow{4}{*}{\texttt{C00\_AR}} & \textit{IFO}\texttt{:GDS-CALIB\_STRAIN\_AR} &  &  & \\
                                      & \textit{IFO}\texttt{:GDS-CALIB\_STRAIN\_NOLINES\_AR} & \checkmark  &  & \\
                                      & \textit{IFO}\texttt{:GDS-CALIB\_STRAIN\_CLEAN\_AR} & \checkmark & \checkmark & \\
                                      & \textit{IFO}\texttt{:GDS-GATED\_STRAIN\_AR} & \checkmark & & \checkmark \\ \hline

    \multirow{3}{*}{\texttt{AR01}} & \textit{IFO}\texttt{:DCS-CALIB\_STRAIN\_AR01} &  &  & \\
                                   & \textit{IFO}\texttt{:DCS-CALIB\_STRAIN\_NOLINES\_AR01} & \checkmark  &  & \\
                                   & \textit{IFO}\texttt{:DCS-CALIB\_STRAIN\_CLEAN\_AR01} & \checkmark & \checkmark & \\ \hline

    \multirow{1}{*}{\texttt{AR1U02}} & \texttt{V1:Hrec\_hoftRepro1AR\_16384Hz} & \checkmark & \checkmark & \\ \hline
  \end{tabular}
\end{table*}

The O4b alternate strain data sets include several versions of the
strain channel for each LIGO instrument, which are
summarized in Table \ref{tab:o4archannels} and described below.

For \texttt{C00\_AR} frames the strain channels are
(\textit{IFO} corresponds to H1 for Hanford or L1 for Livingston):

\begin{itemize}

\item \textit{IFO}\texttt{:GDS-CALIB\_STRAIN\_AR}  This is the calibrated strain
  data with no noise subtraction applied.

\item \textit{IFO}\texttt{:GDS-CALIB\_STRAIN\_NOLINES\_AR} This is the
  calibrated strain data after removing narrowband noise.
  This channel removes the calibration lines, \mainPowerFUS{} power-mains line, and harmonics of the power-mains line as
  described in \citet{vietsdcc}.

\item \textit{IFO}\texttt{:GDS-CALIB\_STRAIN\_CLEAN\_AR} This is the calibrated strain data, after subtracting
  both narrowband and broadband noise.
  This applies additional broadband noise subtraction on to the
  \textit{IFO}\texttt{:GDS-CALIB\_STRAIN\_NOLINES\_AR} channel, using the method described in \citet{2020PhRvD.101d2003V}.
  This is the recommended channel for transient searches, and it corresponds to the channel name
  \texttt{DCS-CALIB\_STRAIN\_CLEAN\_SUB60HZ\_C01\_AR} in O3 \citep{2023ApJS..267...29A}.

\item \textit{IFO}\texttt{:GDS-GATED\_STRAIN\_AR} This is the \textit{IFO}\texttt{:GDS-CALIB\_STRAIN\_NOLINES\_AR} channel
  where times with loud glitches are replaced with zeros by applying a window function around short
  segments of elevated noise \citep{gating, 2016CQGra..33u5004U}.
  This channel was not used for analyses by the LVK.
  An improved version of the algorithm, as used for several O4 CW searches, is described in \citet{T2400003}.

\end{itemize}

Channels in \texttt{AR01} frames follow the same conventions
(replacing \texttt{GDS} by \texttt{DCS} and \texttt{AR} by \texttt{AR01})
with the exception that the glitch removal procedure was not applied
so the corresponding channel doesn't exist.

Finally, in addition to the strain channels, the O4b \texttt{ANALYSIS\_READY} frames
also include a number of auxiliary channels that record data-quality information
and calibration parameters.

Virgo \texttt{ANALYSIS\_READY} frames were produced in chunks of about one month.
The content of the \texttt{ANALYSIS\_READY} frames was only C00 $h(t)$ (produced online)
or C01 $h(t)$ for the first month of O4b where an offline reprocessing was needed.
Whatever its origin (C00 or C01), the name of the Virgo strain channel is \texttt{V1:Hrec\_hoftRepro1AR\_16384Hz}.

The same \texttt{ANALYSIS\_READY} frames also contain the state vector which provides the status of the Virgo interferometer.
The name of this channel is \texttt{V1:DQ\_ANALYSIS\_STATE\_VECTOR} and its value is 4095 when the data quality is optimal
(interferometer is in science mode, $h(t)$ reconstruction is correct and \texttt{CAT1} data-quality flag is not activated).
By definition of the \texttt{ANALYSIS\_READY} frames, $h(t)$ is zero each time the state vector value is not 4095.

Finally, the Virgo \texttt{ANALYSIS\_READY} frames contain six other auxiliary channels present only in the frames whose GPS times are a multiple of 1000.
Those channels provide the frequency-dependent residual bias modulus and phase and the frequency-dependent uncertainties on $h(t)$ modulus and phase.
The prefix of the name of those channels is \texttt{V1:Hrec\_hoftRepro1AR\_U02} and the rest of the channel name is used to specify whether the channel refers to the modulus or the phase (see also \citet{Virgonote_ARframes}).
Details about uncertainty in the calibration for LIGO are provided in separate files as documented in the GWOSC website.\footnote{O4 Data Set Technical Details, \url{https://gwosc.org/O4/o4_details/}}

The alternate strain releases are not currently available for direct
download from the GWOSC website; instead, they are available via OSDF and
NDS2, as discussed earlier in Section \ref{sec:data}.

\section{Auxiliary Channels} \label{sec:auxc}

In addition to the strain time series, LIGO records around \num{200000}
\textit{auxiliary channels} for each instrument.  Auxiliary channels
are stored as time series data and record the state of the instrument,
the local environment, and a number of digital-filter settings.
Auxiliary channels include readings from seismometers, accelerometers,
photodiodes, magnetometers, wind monitors, and microphones
\citep{2021CQGra..38n5001N,2024arXiv241204638H}.

The O3, O4a and O4b data releases include the set of LIGO auxiliary channels
used for noise subtraction and to produce data-quality flags.
These are the ways that auxiliary channels directly contribute
to the GWTC-4.0 and GWTC-5.0 analyses.
For O3, the auxiliary data release includes about 40 channels for each LIGO detector, 
corresponding to roughly 80 channels in total. For the O4a and O4b runs,
the release includes 25 channels total across both interferometers.
The online documentation includes the name of each available
channel along with the associated sampling rate and a short description
of the channel content.\footnote{See
  \url{https://gwosc.org/O3/auxiliary/} for O3 and
  \url{https://gwosc.org/O4/auxiliary/} for O4a and O4b.}

Auxiliary channels should
be used with caution, as the data are not
necessarily calibrated and may include times when the data values
are missing or corrupt.  For example, sensors sometimes break or
become disconnected leading to periods of corrupted data.
Auxiliary data are available at all times when the detector is
in \verb=ANALYSIS_READY= with the exact times marked in segment lists
that are available with the data release documentation.

In addition to channels used for noise subtraction and data-quality flags,
some auxiliary data have been released on request and are now available on the
GWOSC website.
For O3, these releases include a data set to support a study of machine learning
applied to detector characterization \citep{Gurav:2024tkl}
and a study of noise-subtraction methods \citep{zackaydcc}.

Auxiliary-channel data cannot currently be downloaded directly through
the GWOSC website or the API.  Instead, auxiliary-channel data are available
through the NDS2 and OSDF interfaces described in Section \ref{sec:data}.
 \label{auxrelease}

\section{Event Portal}
\label{sec:portal}

The GWOSC Event Portal provides an online interface to a database of published GW transient 
events.\footnote{GWOSC Event Portal, \url{https://gwosc.org/eventapi}} These events are 
short-duration GW signals identified in the strain data by searches. They are typically 
attributed to a signal consistent with a CBC source. Not all events in the Event Portal will 
be from CBCs, some may be noise with an instrumental origin.

Users of the GWOSC will find an array of data products associated with most events available in the 
Event Portal, including strain data, segment lists, detection confidence, astrophysical source parameters,
and documentation. In most cases, the database also includes links to additional analysis 
products, such as posterior samples describing source properties and source localizations.

The Event Portal may be accessed as browsable \texttt{HTML} webpages or via a REST API. A Python 
client, named \verb=gwosc=, can be used to query the API and download 
content.\footnote{GWOSC Client API, \url{https://pypi.org/project/gwosc/}} Documentation of the 
Event Portal and REST API are available on the GWOSC website.

\subsection{User Interfaces to the Data}

GWTC catalogs are a cumulative collection of events. In the Event Portal, \emph{releases} represent sets of 
new events published in specified LVK publications, such as the new CBC candidates added for GWTC-5.0 \citep{GWTC:Introduction, GWTC:Methods, GWTC:Results}, 
or a collection of events that are related in some other way, such as all the GW transients published in 
stand-alone discovery papers such as those from O3. As a consequence there are a number of 
ways to interact with the event data using the Event Portal.

The \emph{Release List View} in the Event Portal displays a list of all available releases, summarized in 
Table \ref{tab:epReleases}. Several releases are labeled marginal, i.e., \texttt{GWTC-1-marginal}, 
\texttt{GWTC-2.1-marginal}, \texttt{GWTC-3-marginal}, \texttt{O3\_IMBH\_marginal}, which include candidate 
triggers that have a low significance more consistent with instrumental background than a true GW event. 

In addition, there is one release, \texttt{GWTC-2.1-auxiliary}, in which revised analyses in 
\texttt{GWTC-2.1} \citep{2024PhRvD.109b2001A} resulted in the demotion of events that were previously 
identified in \texttt{GWTC-2} \citep{2021PhRvX..11b1053A} as statistically significant. These are 
included separately for completeness.

Prior to \texttt{GWTC-4.0}, new events reported on GWOSC matched new events
documented in each publication.
For \texttt{GWTC-4.0} \citep{2025arXiv250818082T}, \texttt{GWTC-4.1}, and \texttt{GWTC-5.0} \citep{GWTC:Results}, 
the events found in the Event Portal satisfy
\begin{equation}
  p_\mathrm{{astro}} > 0.5\ \mathrm{or} \ \mathrm{FAR < \qty{1}{\per\year}},
\end{equation}
which may or may not agree completely with results tabulated in the publications. Here $p_\mathrm{{astro}}$ is 
the probability that a GW candidate is of astrophysical origin \citep{GWTC:Introduction}. 

\texttt{GWTC-4.1} results supersede those in \texttt{GWTC-4.0} and are identified as \texttt{GWTC-4.1}. 
However, as \texttt{GWTC} is a cumulative catalog, all the results labeled \texttt{GWTC-4.1} are included in 
the \texttt{GWTC-5.0} catalog by definition \citep{GWTC:Results}.

Detailed documentation and references to publications are available for each release on the GWOSC website.

\begin{table*}[t]
\centering
\caption{
  List of releases currently available under the GWOSC Event Portal.\footnote{GWOSC Event Portal, \url{https://gwosc.org/eventapi/html}}
}
\label{tab:epReleases}
\begin{tabular}{ll}
\hline
Release Name & Notes \\
\hline
\hline
\texttt{ GWTC } & Cumulative set of GW transients maintained by the LVK \citep{GWTC:Introduction} \\
\texttt{ Initial\_LIGO\_Virgo } & Event release from initial LIGO and Virgo, 2005--2010 \citep{2012PhRvD..85h2002A} \\
\texttt{ O1\_O2-Preliminary } & Notable events in O1 and O2 published prior to GWTC-1.0 release \citep{2019PhRvX...9c1040A} \\
\texttt{ GWTC-1-confident } & Confident detections from the O1 and O2 runs \citep{2019PhRvX...9c1040A} \\
\texttt{ GWTC-1-marginal } & Marginal candidates from O1 and O2 runs \citep{2019PhRvX...9c1040A} \\
\texttt{ GWTC-2 } & Confident detections from the O3a run \citep{2021PhRvX..11b1053A} \\
\texttt{ GWTC-2.1-confident } & Confident detections from O3a run based on a revised analysis \citep{2024PhRvD.109b2001A} \\
\texttt{ GWTC-2.1-marginal } & Marginal candidates from O3a run based on a revised analysis \citep{2024PhRvD.109b2001A} \\
\texttt{ GWTC-2.1-auxiliary } & Candidates from GWTC-2.0 that were demoted in GWTC-2.1 \citep{2021PhRvX..11b1053A} \\
\texttt{ GWTC-3-confident } & Confident detection from O3b run \citep{2023PhRvX..13d1039A} \\
\texttt{ GWTC-3-marginal } & Marginal candidates from O3b run \citep{2023PhRvX..13d1039A} \\
\texttt{ IAS-O3a } & Events from O3a as described in a community catalog \citep{2022PhRvD.106d3009O} \\
\texttt{ O3\_Discovery\_Papers } & Notable events in O3 run published independently \citep{2020ApJ...892L...3A, 2020PhRvD.102d3015A, 2020ApJ...900L..13A, 2020ApJ...896L..44A, 2021ApJ...915L...5A} \\
\texttt{ O3\_IMBH\_marginal } & Marginal intermediate-mass black hole candidates from O3 run {\catcode`\&=12\cite{2022A&A...659A..84A}}\\
\texttt{ O4\_Discovery\_Papers } & Notable events in O4 run published independently \citep{2024ApJ...970L..34A, 2025arXiv250907348T, 2025ApJ...993L..25A, 2025ApJ...993L..21A,2025PhRvL.135k1403A, 2026arXiv260511703T}\\
\texttt{ GWTC-4.0 } & Data release for O4a events \citep{2025arXiv250818082T} \\
\texttt{ GWTC-4.1 } & Updated data release for O4a based on a revised analysis \citep{GWTC:Results} \\
\texttt{ GWTC-5.0 } & Data release for O4b events \citep{GWTC:Results} \\
\hline
\end{tabular}
\end{table*}

By selecting one of the releases, an \emph{Event List} table is generated showing all events in the release, along 
with source parameters and credible intervals where available. The Event List view includes estimates of a number of 
physical parameters, such as mass, distance, and spin, as well as search-pipeline outputs such as $\mathrm{SNR}$ and $\mathrm{FAR}$. If more 
than one parameter-estimation analysis is available for any event, then values from the analysis marked as \texttt{Default PE} 
for the specific event are displayed in the Event List to indicate which of the multiple results are captured in the 
table. Additionally, the lowest $\mathrm{FAR}$ and highest probability of astrophysical origin $p_\mathrm{{astro}}$ 
from all available search results are included in the Event List table.  

A special case of the Event List view is the \texttt{GWTC} release, which always corresponds to an electronic version of 
the latest cumulative GWTC \citep[currently GWTC-5.0;][]{GWTC:Introduction, GWTC:Methods, GWTC:Results}. Selecting the GWTC 
release returns this set of events as an Event List. It includes confidently detected events from multiple data releases. 

Event Lists are also available to browse specific events using customized queries. Event Lists may include more than 
one version of an event, as some events have appeared in multiple publications, or have been re-analyzed with newer 
code or models. The \emph{Query Page} presents the user with a form that allows selection of events based on name, 
membership in particular releases, various ranges of masses, distance, redshift, effective inspiral spin ($\chi_\mathrm{eff}$), 
$p_\mathrm{astro}$, $\mathrm{SNR}$, $\mathrm{FAR}$, $\mathrm{UTC}$ and $\mathrm{GPS}$ times. The query also allows 
specification of the output format to return, with support for \texttt{HTML}, \texttt{JSON}, \texttt{CSV} and \texttt{ASCII}.

The \emph{Single Event View} shows key parameters for an individual event in the Event Portal. The view can include 
sets of parameters from different pipelines, including search-pipeline outputs such as $\mathrm{SNR}$ and $\mathrm{FAR}$, and inferred 
source properties such as masses and spins. The Single Event View also includes images showing a 
spectrogram \citep{2005PhDT.......307C} of the strain data for each detector at the time of the event, and links to
download the associated strain time series in multiple file formats, as described in Section \ref{sec:data}. The 
Single Event View also provides links to documentation, low-latency information in the Gravitational-Wave Candidate 
Event Database (GraceDB),\footnote{GraceDB website, \url{https://gracedb.ligo.org/}} General Coordinates Network 
(GCN) Circulars and Notices, and links to segment lists for times around the event (see Section \ref{sec:DqInj} for 
details). A new feature found on these pages is the Event Viewer which leads to a web app that generates plots of 
waveforms, source parameters and sky localization for each GW event.\footnote{GW Event Viewer, \url{https://peviewer.igwn.org}}

\subsection{Parameter-Estimation Results}
\label{sec:EPpe}

Most events in the Event Portal include one or more sets of credible intervals drawn from the publication associated 
with the event. The naming of parameters is based on the LVK standard names convention.\footnote{Parameter Estimation 
Name Standard, \url{https://lscsoft.docs.ligo.org/pesummary/stable/gw/parameters.html}} Many entries include multiple 
sets of values attributed to different priors, waveforms, or analysis changes \citep{GWTC:Methods}. In some cases, a 
publication will present credible intervals describing samples combined from several analyses using different waveform 
models. The results for any individual parameter are expressed as the median value with its $90\%$ symmetric credible 
interval.  Details regarding how the posterior samples were constructed can be found in the publication from which the 
event is released or in the supplemental releases linked from the Single Event View.  

\subsection{Supplemental Data Releases}
\label{sec:EPsupdr}

For most events, the Single Event View includes links to supplemental data. These data vary between releases, but in 
general will include links to the parameter-estimation posterior samples, source localization, and additional resources. 
Supplemental data releases are published either in the LIGO Document Control Center\footnote{Public LIGO Document Control 
Center, \url{https://dcc.ligo.org/public/}} or under the LVK Zenodo Community.\footnote{LVK Zenodo 
Data Releases, \url{https://zenodo.org/communities/ligo-virgo-kagra/}} In addition, GWOSC maintains snapshots of the 
Event Portal's version history as a data set available on 
Zenodo.\footnote{GWOSC Zenodo Event Portal Snapshots, \url{http://doi.org/10.1088/1361-6382/ad0b9b}} Each snapshot consists 
of \texttt{JSON} files downloaded from the Event Portal with data for each event.

\subsection{Community Catalogs}
\label{sec:EPcommcat}

The Event Portal has recently been extended to provide GW events from catalog authors outside of the LVK, which are 
collectively organized as \emph{Community Catalogs}. The presentation of results for Community Catalogs is identical 
to that available with any LVK release of events, adhering to the website layout and search capabilities discussed 
previously. This is achieved by standardizing on a \texttt{JSON} schema file format for GWOSC staff to use to ingest 
search and parameter-estimation values. The details for this standard \texttt{JSON} are publicly available in 
GitHub,\footnote{GWOSC Github Community Catalog Schema, \url{https://github.com/gwosc-tutorial/gwosc-catalog}} along 
with useful test scripts, notebooks, and documentation. The criteria for authors outside the LVK Collaboration to 
have their catalogs included in GWOSC is documented in the \emph{GWOSC Community Catalogs Guidelines}.\footnote{GWOSC Community Catalogs Guidelines, \url{https://dcc.ligo.org/LIGO-M2500012/public}}

\section{Summary}

Data from the LVK network of observatories are publicly available through the GWOSC website at \href{https://gwosc.org}{gwosc.org}, 
and the latest update includes O4b data from LIGO and Virgo.
This paper can serve as a practical guide for users interested in analyzing this set of open data.
It provides important details about calibration and data quality, explains the structure of the data sets
available online, and offers instructions for using the Event Portal, which gives access to the list of
astrophysical sources identified by the LVK.

Making these resources available maximizes the scientific potential of the data set.
The O4b data contain a wealth of compact object mergers, 
many of which are listed in version 5.0 of the Gravitational-Wave Transient Catalog \citep{GWTC:Results}.
By releasing the data publicly, the community is empowered to further explore and discover new insights.

The present release covers the second part of O4. The third part is
planned for release in December 2026 \citep{LIGO-DMP}.
With the completion of the O4 data set, we anticipate even more opportunities for discovery.

\section*{Acknowledgements}

\newif\ifcoreonly
\newif\ifkagra
\newif\ifheader
\coreonlytrue  \kagratrue  \headertrue
This material is based upon work supported by NSF's LIGO Laboratory, which is a
major facility fully funded by the National Science Foundation.
The authors also gratefully acknowledge the support of
the Science and Technology Facilities Council (STFC) of the
United Kingdom, the Max-Planck-Society (MPS), and the State of
Niedersachsen/Germany for support of the construction of Advanced LIGO 
and construction and operation of the GEO\,600 detector. 
Additional support for Advanced LIGO was provided by the Australian Research Council.
The authors gratefully acknowledge the Italian Istituto Nazionale di Fisica Nucleare (INFN),  
the French Centre National de la Recherche Scientifique (CNRS) and
the Netherlands Organization for Scientific Research (NWO)
for the construction and operation of the Virgo detector
and the creation and support  of the EGO consortium. 
The authors also gratefully acknowledge research support from these agencies as well as by 
the Council of Scientific and Industrial Research of India, 
the Department of Science and Technology, India,
the Science \& Engineering Research Board (SERB), India,
the Ministry of Human Resource Development, India,
the Spanish Agencia Estatal de Investigaci\'on (AEI),
the Spanish Ministerio de Ciencia, Innovaci\'on y Universidades,
the European Union NextGenerationEU/PRTR (PRTR-C17.I1),
the ICSC - CentroNazionale di Ricerca in High Performance Computing, Big Data
and Quantum Computing, funded by the European Union NextGenerationEU,
the Comunitat Auton\`oma de les Illes Balears through the Conselleria d'Educaci\'o i Universitats,
the Conselleria d'Innovaci\'o, Universitats, Ci\`encia i Societat Digital de la Generalitat Valenciana and
the CERCA Programme Generalitat de Catalunya, Spain,
the Polish National Agency for Academic Exchange,
the National Science Centre of Poland and the European Union - European Regional
Development Fund;
the Foundation for Polish Science (FNP),
the Polish Ministry of Science and Higher Education,
the Swiss National Science Foundation (SNSF),
the Russian Science Foundation,
the European Commission,
the European Social Funds (ESF),
the European Regional Development Funds (ERDF),
the Royal Society, 
the Scottish Funding Council, 
the Scottish Universities Physics Alliance, 
the Hungarian Scientific Research Fund (OTKA),
the French Lyon Institute of Origins (LIO),
the Belgian Fonds de la Recherche Scientifique (FRS-FNRS), 
Actions de Recherche Concert\'ees (ARC) and
Fonds Wetenschappelijk Onderzoek - Vlaanderen (FWO), Belgium,
the Paris \^{I}le-de-France Region, 
the National Research, Development and Innovation Office of Hungary (NKFIH), 
the National Research Foundation of Korea,
the Natural Sciences and Engineering Research Council of Canada (NSERC),
the Canadian Foundation for Innovation (CFI),
the Brazilian Ministry of Science, Technology, and Innovations,
the International Center for Theoretical Physics South American Institute for Fundamental Research (ICTP-SAIFR), 
the Research Grants Council of Hong Kong,
the National Natural Science Foundation of China (NSFC),
the Israel Science Foundation (ISF),
the US-Israel Binational Science Fund (BSF),
the Leverhulme Trust, 
the Research Corporation,
the National Science and Technology Council (NSTC), Taiwan,
the United States Department of Energy,
and
the Kavli Foundation.
The authors gratefully acknowledge the support of the NSF, STFC, INFN and CNRS for provision of computational resources.

This work was supported by MEXT,
the JSPS Leading-edge Research Infrastructure Program,
JSPS Grant-in-Aid for Specially Promoted Research 26000005,
JSPS Grant-in-Aid for Scientific Research on Innovative Areas 2402: 24103006,
24103005, and 2905: JP17H06358, JP17H06361 and JP17H06364,
JSPS Core-to-Core Program A.\ Advanced Research Networks,
JSPS Grants-in-Aid for Scientific Research (S) 17H06133 and 20H05639,
JSPS Grant-in-Aid for Transformative Research Areas (A) 20A203: JP20H05854,
the joint research program of the Institute for Cosmic Ray Research,
University of Tokyo,
the National Research Foundation (NRF),
the Computing Infrastructure Project of the Global Science experimental Data hub
Center (GSDC) at KISTI,
the Korea Astronomy and Space Science Institute (KASI),
the Ministry of Science and ICT (MSIT) in Korea,
Academia Sinica (AS),
the AS Grid Center (ASGC) and the National Science and Technology Council (NSTC)
in Taiwan under grants including the Science Vanguard Research Program,
the Advanced Technology Center (ATC) of NAOJ, 
the Mechanical Engineering Center of KEK
and Vietnam National Foundation for Science and Technology Development 
(NAFOSTED) 103.01-2025.147.

Additional acknowledgements for support of individual authors may be found in the following document: \\
\texttt{https://dcc.ligo.org/LIGO-M2300033/public}.
For the purpose of open access, the authors have applied a Creative Commons Attribution (CC BY)
license to any Author Accepted Manuscript version arising.
We request that citations to this article use 'A. G. Abac {\it et al.} (LIGO-Virgo-KAGRA Collaboration), ...' or similar phrasing, depending on journal convention.

\textit{Software:} This software infrastructure used to publish data products onto the GWOSC website
made significant use of the IGWN Software Environment \citep{igwn-env} curated by the
LVK to provide a reproducible computing environment built around tools
needed for GW data analysis.
Plots were prepared with \texttt{Matplotlib}~\citep{2007CSE.....9...90H}.

\section*{Data Availability}

All data products described in this work are publicly available on the
Gravitational Wave Open Science Center,\footnote{GWOSC Home Page: \url{https://gwosc.org}}
the Zenodo LVK community page,\footnote{Zenodo LVK Community Page: \url{https://zenodo.org/communities/ligo-virgo-kagra/}}
and through the OSDF and NDS2 data repositories, as described in the previous sections of this paper.
Data products are provided under a Creative Commons Attribution 4.0 International
license.\footnote{CC BY 4.0 license: \url{https://creativecommons.org/licenses/by/4.0/}}

\appendix
\section{Additional Segments in the O4a Strain-Data Release}
\label{sec:O4a_bulk}

The O4a run started on 24 May 2023 at 15:00 UTC. 
However, a few segments of data from the engineering run just prior to the start
of the run were used in the search for GW emitted from SN 2023ixf \citep{2025ApJ...985..183A}
and for the neutron star--black hole binary candidate GW230518\_125908 \citep{GWTC:Results},
and they have been added to this release.
These additional data segments span between times 15 May 2023 14:13:22 UTC (GPS 1368195220)
and 19 May 2023 17:31:33 UTC (GPS 1368552711) when both LHO and LLO were in observing mode.
These segments have a total duration of 0.8 days.
Including this addition, the total amount of observing time in O4a is 160.8
days for LHO and 164.3 days for LLO (to compare with Table~\ref{dutycycle_O4}, 0.8
days have to be added to the coincident times).

Using this total amount of time as reference, the percentage of the observing time for which the data quality pass each category described in Table~\ref{tab:dq}
in O4a can be calculated.
Since these percentages are close to
100\%, Table~\ref{tab:dqO4a} summarizes the percentages of failing times.

\begin{table}[ht]
\centering
\caption{Percentage of the observing time in O4a, including the extra 0.8 days of engineering run data, for which data quality fail the categories described in Table~\ref{tab:dq}. }
\vspace{10pt}
\label{tab:dqO4a}
\begin{tabular}{lll}
\hline
Short name	&	Percentage in LHO				&	Percentage in LLO	\\
\hline
\hline
\texttt{DATA}			&	0.0971\% 			&	0.0059\% \\
\texttt{CBC\_CAT1} 		&	0.0971\%			&	0.0059\%\\
\texttt{CBC\_CAT2} 		&	0.0971\%			&	0.0059\%\\
\texttt{CBC\_CAT3} 		&	0.0971\%			&	0.0059\%\\
\texttt{BURST\_CAT1} 	&	0.0971\%			&	0.0059\%\\
\texttt{BURST\_CAT2} 	&	0.2242\%		 	&	0.1188\%\\
\texttt{BURST\_CAT3} 	&	0.2242\% 			& 	0.1188\% \\
\texttt{STOCH\_CAT1} 	&	0.0979\% 			&	0.0065\%\\
\texttt{CW\_CAT1} 		&	0.0971\% 			&	0.0059\%\\
\hline
\end{tabular}
\end{table}  

\bibliography{}

\end{document}